\documentclass[twocolumn]{emulateapj}
\usepackage{subfigure}
\usepackage{amsmath}
\begin{document}
\title{A semi-analytical model for exploring Galilean satellites formation from a massive disk}
\author{Yamila Miguel}
\affil{Laboratoire Lagrange, UMR 7293, Universit\'e de Nice-Sophia Antipolis, CNRS, Observatoire de la C\^ote d’Azur, Blvd de l'Observatoire, CS 34229, 06304 Nice cedex 4, France\\
Max Planck Institut f\"ur Astronomie, K\"onigstuhl 17, 69117, Heidelberg, Germany}
\email{yamila.miguel@oca.eu}
\author{Shigeru Ida} 
\affil{Earth-Life Science Institute, Tokyo Institute of Technology, Meguro-ku, Tokyo 152-8550, Japan}

\begin{abstract}
A better knowledge of Jovian satellites' origins will bring light on the environment that surrounded Jupiter during its formation and can help us to understand the characteristics of this unique satellite system. We developed a semi-analytical model to investigate Jupiter's regular satellite formation and present the results of our population synthesis calculations. We performed simulations adopting a massive, static, low-viscosity circumplanetary disk model, in agreement with a current study of magnetorotational instability in a circum-planetary disk. We find that the high gas density leads to very rapid migration of satellitesimals due to gas drag and type II migration of satellites in a faster disk-dominated mode. A large concentration of solids, large building blocks and longer type II migration time-scales favor formation and survival of large satellites. However, bodies as massive as Ganymede and those located far away from Jupiter, such as Callisto, are difficult to form with this scenario.
\end{abstract}

\keywords{Jupiter, satellites - satellites, formation - satellites, general}

\section{Introduction}
\label{intro}
                                                    
The conditions that lead to Galilean satellites formation are important for understanding their properties and for putting additional constrains on Jupiter's formation history. Their almost circular, low inclined orbits lead to the general agreement that these satellites formed from a circumplanetary disk that surrounded Jupiter during its formation \citep{ls82}. This disk can be described by two different models: the gas-starved disk model \citep{cw02,cw06,cw09} and the minimum mass model \citep{ls82,mea03a,meb03b,e09}. 

In the gas-starved disk model Jovian satellites grow at the same time as Jupiter in a viscously evolving disk, where the inflow of material attracted by Jupiter provides a continuos source for satellites formation, maintaining a low density in the disk at all times \citep{cw02,al05,cw06,cw09}. This model was used for different authors to explore regular satellites formation in the Solar System adopting different approaches. \citet{og12} performed N-body simulations for satellite formation, introducing the idea of adding an inner cavity in the disk. Because the satellite's migration is halted at the inner disk edge
and the innermost satellite suffers a strong torque called "eccentricity trap " from the disk \citep{og10}, a satellite system trapped in mutual 2:1 resonances were successfully formed. They also argued the robustness of the resonant trapping. \citet{sa10} developed a semi-analytical model based on the work by \citet{il04,il08}, using the gas-starved disk model. In their simulations, systems of 3-5 similar mass satellites trapped in resonances like Galilean satellites are formed if the inner cavity and rapid removal of the circumplanetary protosatellite disk due to gap formation in a circumstellar protoplanetary disk, which may correspond to conditions of proto-Jupiter, are assumed. On the other hand, systems dominated by a single large satellite in relatively outer region like Saturnian system are formed if no inner cavity and gradual dissipation of the circumplanetary disk, which may correspond to
conditions of proto-Saturn,  are assumed. Note that we need more detailed studies on the inner cavity and removal of protosatellite disks to justify their scenario. The gas-starved disk model was also used for simulations of satellite formation outside the Solar System \citep{he15a,he15b}.

On the other hand, according to the minimum mass model, the infall of gas into Jupiter through the Lagrange points at the last stages of its formation allows the accumulation of a massive, static disk with low viscosity that could give birth to Jovian regular satellites. This disk had the solid surface density necessary to form the observed Galilean satellites with their corresponding rock/ice mass ratio. The main difference between this model and the gas-starved disk model is that in this case the disk is driven by a weak turbulent regime, which allows the accumulation of a massive disk, changing the conditions for Galilean satellites formation. 

A recent paper by \citet{fu14} showed that MRI is unlikely to be well developed in circumplanetary disks, because magnetic diffusion is very fast in the disks that are much more compact than circumstellar disks. \citet{tu14} also showed that most parts of the disk are likely to be MRI dead. If MRI does not occur, the resultant weak viscosity should lead to a large accumulation of gas, which may be rather more consistent with a minimum mass disk model than a gas-starved disk.

Since at present it is not clear which disk model is more appropriate for the Jovian satellite disk, we need to investigate accretion and orbital evolution of large satellites in circumplanetary disks similar to the massive minimum mass disk model, as well as those with the gas-starved disk model. The latter model has already been explored by N-body simulations \citep{cw09,og12} and by population synthesis calculations \citep{sa10}, while the minimum mass model has not been further explored. With this motivation, we developed a semi-analytical model to investigate Jupiter regular satellites' formation following the minimum mass Jovian-subnebula description. Our model is based on papers by \citet{il04,il08,il10} and \citet{mb08,mb09,mb10}. 

During the formation process, satellitesimals migrate due to gas drag and proto-satellites migrate due to disk-planet interaction (see section \ref{a-evo}). The minimum-mass model neglects orbital migration of bodies and therefore, the final distribution of satellites does not necessarily follow the mass distribution of the minimum-mass model. In this paper, we explore different parameters, investigating how the formed satellite distribution is affected by the migration and resonant trapping between satellites, towards a better understanding of the initial conditions that lead to Galilean satellites formation. 

\section{The model}

Our calculations start when the gas infall onto Jupiter's circumplanetary disk stops. At this last stage in Jupiter formation, the turbulence in the subnebula decays allowing the formation of Galilean satellites. The current characteristics of Galilean satellites are shown in table \ref{galileanos}. 

\begin{deluxetable*}{lccc} 
\tablecolumns{4} 
\tablewidth{0pc} 
\tablecaption{Characteristics of Galilean Satellites \citep{be99}} 
\tablehead{
\colhead{Name} & \colhead{Semimajor axis (R$_{Jup}$)} &\colhead{Radius (Km)} &\colhead{Mass (g) [$M_{Jup}$]}}
\startdata
Io & 5.90 & 1821 & 8.94$\times10^{25}$ $[4.7 \times 10^{-5}]$\\
Europa & 9.94 & 1565 & 4.8$\times10^{25}$ $[2.5 \times 10^{-5}]$\\
Ganymede & 14.99 & 2634 & 1.48$\times10^{26}$ $[7.8 \times 10^{-5}]$\\
Callisto & 26.37 & 2403 & 1.07$\times10^{26}$ $[5.6 \times 10^{-5}]$
\enddata 
\label{galileanos}
\end{deluxetable*} 

\subsection{Initial disk structure}\label{disc-1}

The minimum-mass protoplanetary disk model for Jupiter that we consider here is composed of two parts: an optically thick inner region inside Jupiter's centrifugal radius and an outer disk extended up to 150 R$_{Jup}$, near the location of Jupiter's first irregular satellites. To obtain the size of the optically thick inner disk we assume that the gas falling through the Lagrange points conserves circumplanetary angular momentum, $l$. For gas moving on a Keplerian orbit and assuming that Jupiter's atmosphere fills its lobe it is found that
\begin{equation}
l \simeq \frac{1}{4} \Omega R^2_{Hill}
\end{equation}
with $\Omega$ the Kepler frequency of the planet moving around the Sun and $R_{Hill}$ Jupiter's Hill radius \citep{li95}. Using this value for $l$, Jupiter's centrifugal radius ($r_c$) is obtained from equating the gravitational to the centrifugal forces:
\begin{equation}
\frac{l^2}{r_c^3}\simeq \frac{GM_{Jup}}{r_c^2}
\end{equation} 
with $G$ the gravitational constant and $M_{Jup}$ Jupiter's mass, resulting in $r_c\simeq R_{Hill}/48$, approximately 15 Jupiter's radius (R$_{Jup}$) \citep{st86,mea03a}.

Data from gravity measurements by the Galileo mission and interior structure models showed that Galilean satellites have a strong monotonic variation in ice fraction with distance from Jupiter: Io is $100 \%$ rock, Europa is $\sim 90 \%$ rock and $\sim 10 \%$ ice and Ganymede and Callisto are $\sim 50 \%$ rock and $\sim 50 \%$ ice by mass \citep{so02}. In order to satisfy these constrains, we adopted a temperature gradient in the subnebula due to Jupiter's luminosity that is characterized with very high temperatures in the inner disk and much lower, ice condensation temperatures in the outer regions \citep{mea03a}. However, if orbital migration of satellites is taken into account the original disk compositional gradients are smoothed out and it is not easy to reproduce the compositional gradients of Galilean satellites \citep{dw13}. For Io and Europe, tidal heating could also vaporize icy components. We leave the problem of the compositional gradients for future work and are focused on dynamical properties.

\begin{figure}[ht]
\begin{center}
\includegraphics[angle=90,scale=.29]{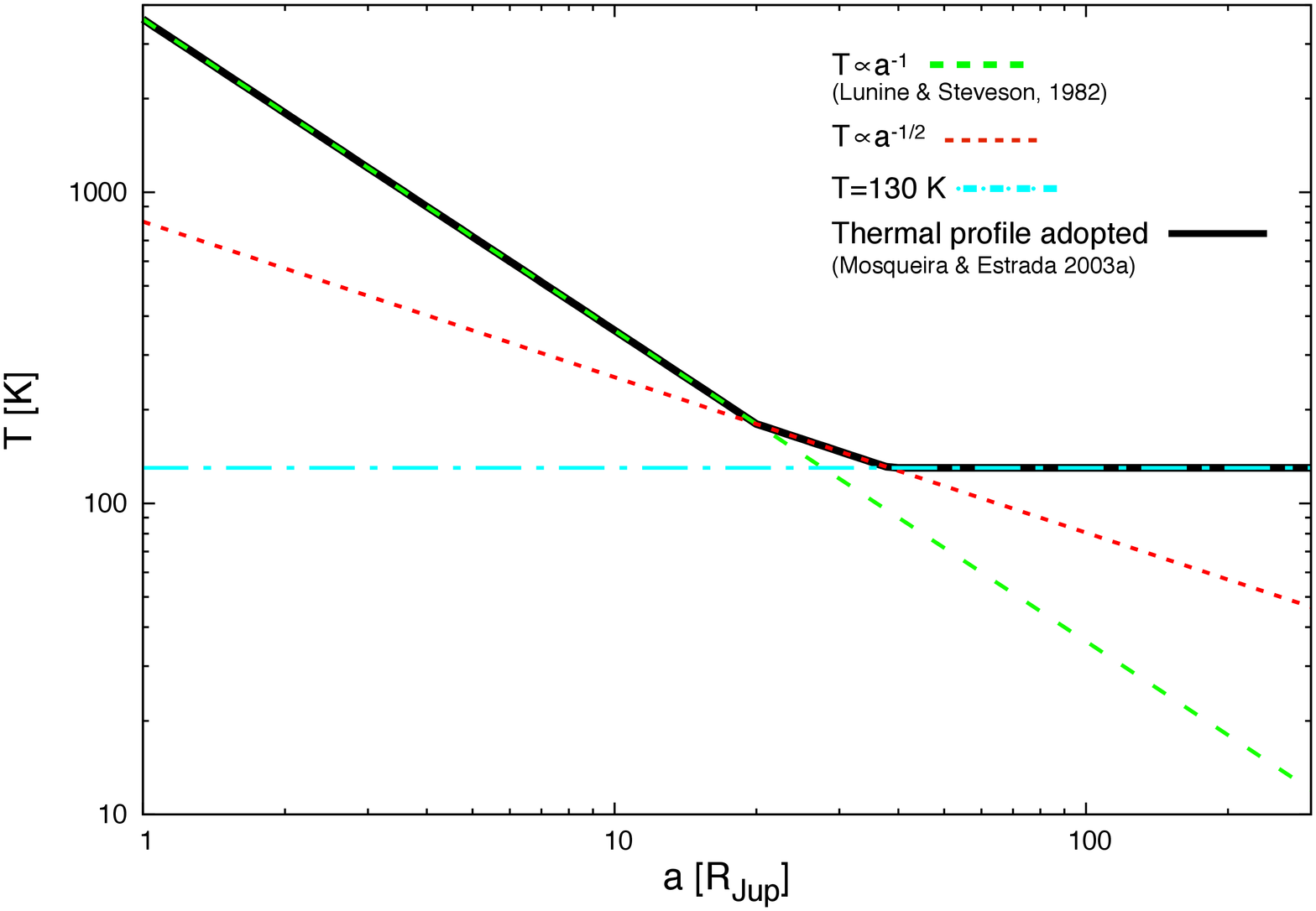}
\caption{Thermal profile of Jupiter's protoplanetary disk adopted in this paper (black solid), which is a combination of a viscous driven thermal profile proportional to $a^{-1}$ in the inner disk (green dashed), an intermediate profile proportional to $a^{-1/2}$ (red dotted) and a constant profile of 130 K in the outer disk (cyan dotted and dashed line).}
\label{temperature}
\end{center}
\end{figure}

Figure \ref{temperature} shows the temperature of the disk as a function of semimajor axis. The thermal structure (black solid line) is separated in three regions. In the inner, optically thick region the temperature is determined by viscous heating, and is characterized by an adiabatic profile $T=3600 \frac{R_{Jup}}{a}$ (green dashed line), where the constant is chosen setting the present radial distance of Ganymede as the ice condensation radius \citep{ls82}. The disk extends up to an outer, optically thin region which has a constant value corresponding to the solar nebula temperature at the location of Jupiter ($\simeq$130 K) (cyan dotted and dashed line). In the intermediate, optically thin region of the disk a radiative profile characterizes the temperature, where the main source of irradiation is due to Jupiter's luminosity and $T \propto a^{-1/2}$ (red dotted line). 

The empirical minimum-mass model adopted in this work to represent Jupiter's nebula is an approximation. In reality, the gas surface density distribution in Jupiter's protosatellite disk should be determined by a balance between the infall and viscous accretion onto the planet, while the solid surface density should be determined by the infall, capture of passing planetesimals and migration due to gas drag. Nevertheless, the large uncertainties in the disk viscosity, infall rate, capture rate of planetesimals, size of satellitesimals and migration of satellites, make the model a useful reference starting point. A highly detailed model of Jupiter's nebula is out of the scope of this paper. In this model, the disk's gas surface density ($\Sigma_g$) is given by equation \ref{gas} \citep{mea03a}:

\begin{equation}\label{gas}
\Sigma_g(a) = \left\{ \begin{array}{ll}
 51\times10^4 \left(\frac{R_{in}}{a}\right)&\textrm{if $a<20R_{Jup}$}\\
\\
 2.92\times10^{22}\left(\frac{a}{R_{Jup}}\right)^{-13}&\textrm{if $20<a<26R_{Jup}$}\\
\\
 0.31\times10^4\left(\frac{R_{out}}{a}\right)&\textrm{if $26< a <150R_{Jup}$}\\
 0  &\textrm{if $a >150R_{Jup}$}\
\end{array} \right.
\end{equation}
with $\Sigma_g(a)$ in $\frac{g}{cm^2}$, $R_{in}=14R_{Jup}$, $R_{out}=87R_{Jup}$ and $a$ is the semimajor axis of satellites' orbit around Jupiter. 

\begin{figure}[ht]
\begin{center}
\includegraphics[angle=90,scale=.28]{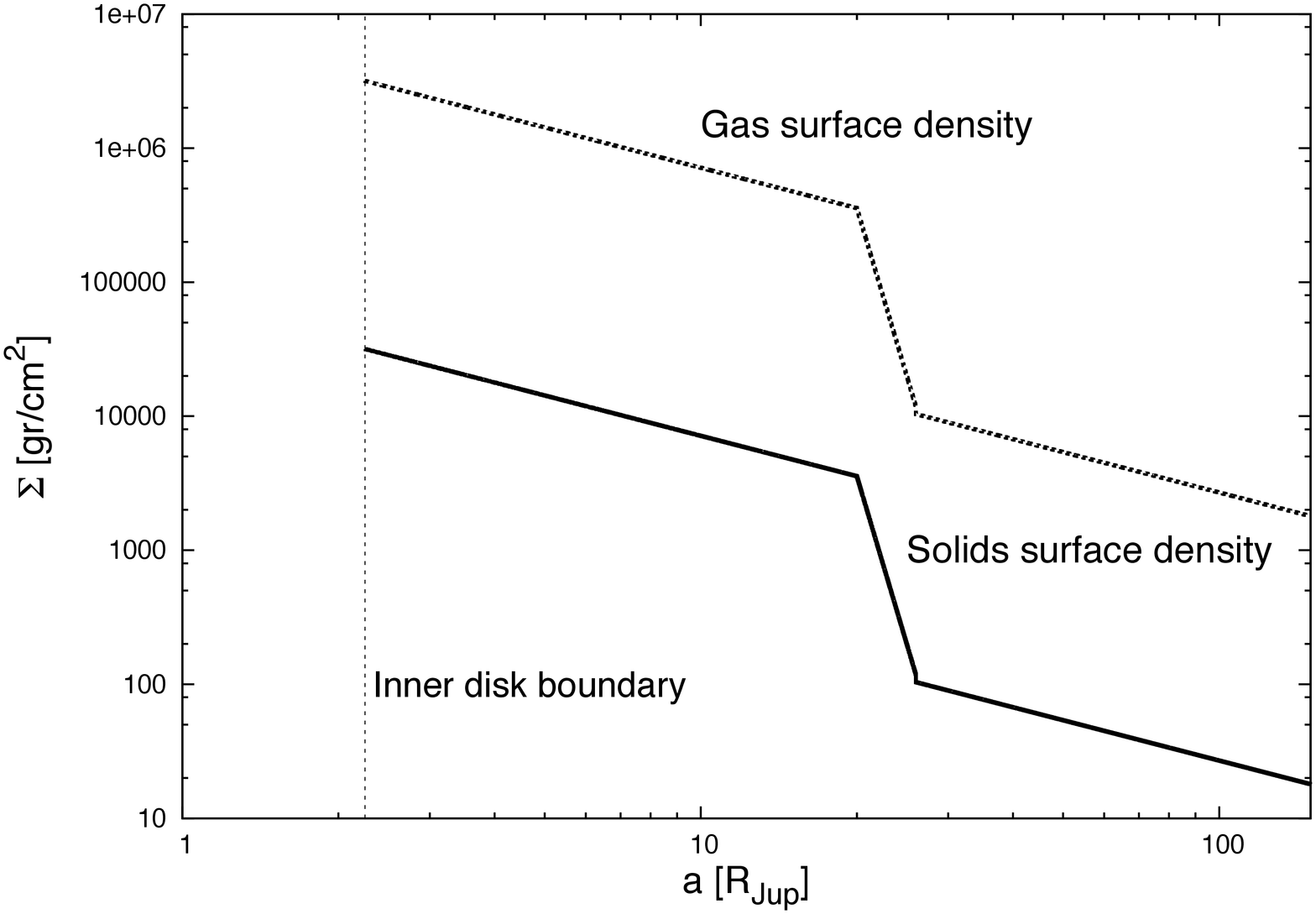}
\caption{Initial gas (dotted) and solid (full line) surface density when the gas to dust ratio adopted is 100. The inner disk cavity is shown  (dashed line).}
\label{disc}
\end{center}
\end{figure}

For the solid disk, we calculate the surface density ($\Sigma_s$) assuming that the gas to dust ratio is equal to 100, the same value as in the solar nebula. Figure \ref{disc} shows the gas and solid disk profiles. Here we neglect an increase in solid surface density due to ice condensation, because uncertainty in the initial solid surface density would be larger than a factor of 2 (the bulk faction of ice in Ganymede and Callisto is estimated as ~1/2).

The total disk masses of gas and solids are $\sim 2\times10^{-2}$ $M_{Jup}$ and $\sim 2\times10^{-4}$ $M_{Jup}$, respectively. 
Satellitesimal distribution can evolve differently than gas distribution and solids can be supplied by ablation of captured heliocentric planetesimals, allowing an enhancement of solids by a factor of $\sim$10 above solar, which is also consistent with the solid enhancement observed in the Jovian atmosphere \citep{e09}. Therefore, we also performed some simulations adopting a gas to dust ratio of 10 ($\Sigma_g$ is reduced or $\Sigma_s$ is enhanced; see section \ref{section-gas}). 

During formation, giant planets experienced despinning from a rapid primordial rotation rate to their present spin. \citet{ts96} studied the angular momentum transfer from Jupiter to the proto-satellites nebula due to the interaction of the planetary magnetic field with this disk. They found that the disk is truncated at the corotation radius where the gaseous disk corotates with the planetary spin. In order to take this effect into account, we introduce an internal cavity in the disk which sets an inner boundary at $\sim 2.25~R_{Jup}$ \citep{sa10}.  
  
\subsection{Satellites' growth}\label{growth}
The process of satellite-embryos formation in circumplanetary disks is not well understood and even formation of planetesimals in circumstellar disks has not been identified yet. Nevertheless, a study by \citet{jo07} propose rapid formation of large planetesimals from meter-sized bodies in a turbulent circumstellar disk. After that, the large planetesimals can grow rapidly to much larger bodies by sweeping up pebbles \citep{ok10,lj12,lj14}. \citet{mea03a} proposed a similar scenario for the formation of satellite-embryos in a circumplanetary disk, where satellites' first stage of growth is due to sweep-up of dust and rubble from the solid disk, which leads to 1000 Km-sized satellite-embryos rapidly. 

Following these ideas, we set satellite-embryos of physical radius of 1000 km and small satellitesimals (of 1, 10, 20 or 30 km, see section \ref{nebula-evolution}) embedded in a gas disk as initial conditions. In the massive disk, satellitesimal random velocities are strongly damped by aerodynamical gas drag and the satellitesimal disk becomes very thin, making accretion a two-dimensional process, with high collisional probability and focussing factors. The characteristic protosatellite accretion time-scale in this shear-dominated regime is given by equation \ref{accretionrate} \citep{ida90,ra04}.

\begin{equation}\label{accretionrate}
\tau_{acc}=\frac{M_{sat}}{\dot{M}_{sat}}
\end{equation}
\begin{equation}\label{solids-accretion}
\begin{split}
\dot{M}_{sat}
\simeq 2R_{sat} \Sigma_s \sqrt{\frac{GM_{sat}}{R_{sat}v^2}} v =\\
2 \left(\frac{R_{sat}}{a}\right)^{\frac{1}{2}} \Sigma_s a^2
\left(\frac{M_{sat}}{M_{Jup}}\right)^{\frac{1}{2}} \frac{2\pi}{T_k}
\end{split}
\end{equation}
where $M_{sat}$ the protosatellite mass, $R_{sat}$ is the protosatellite's physical radius and $T_k$ is the kepler period. The characteristic accretion time-scale for a satellite like Europa  is $9.5$ years and for Ganymede is $19$ years (table \ref{galileanos}). 

\begin{figure}[ht]
\begin{center}
\includegraphics[angle=90,scale=.28]{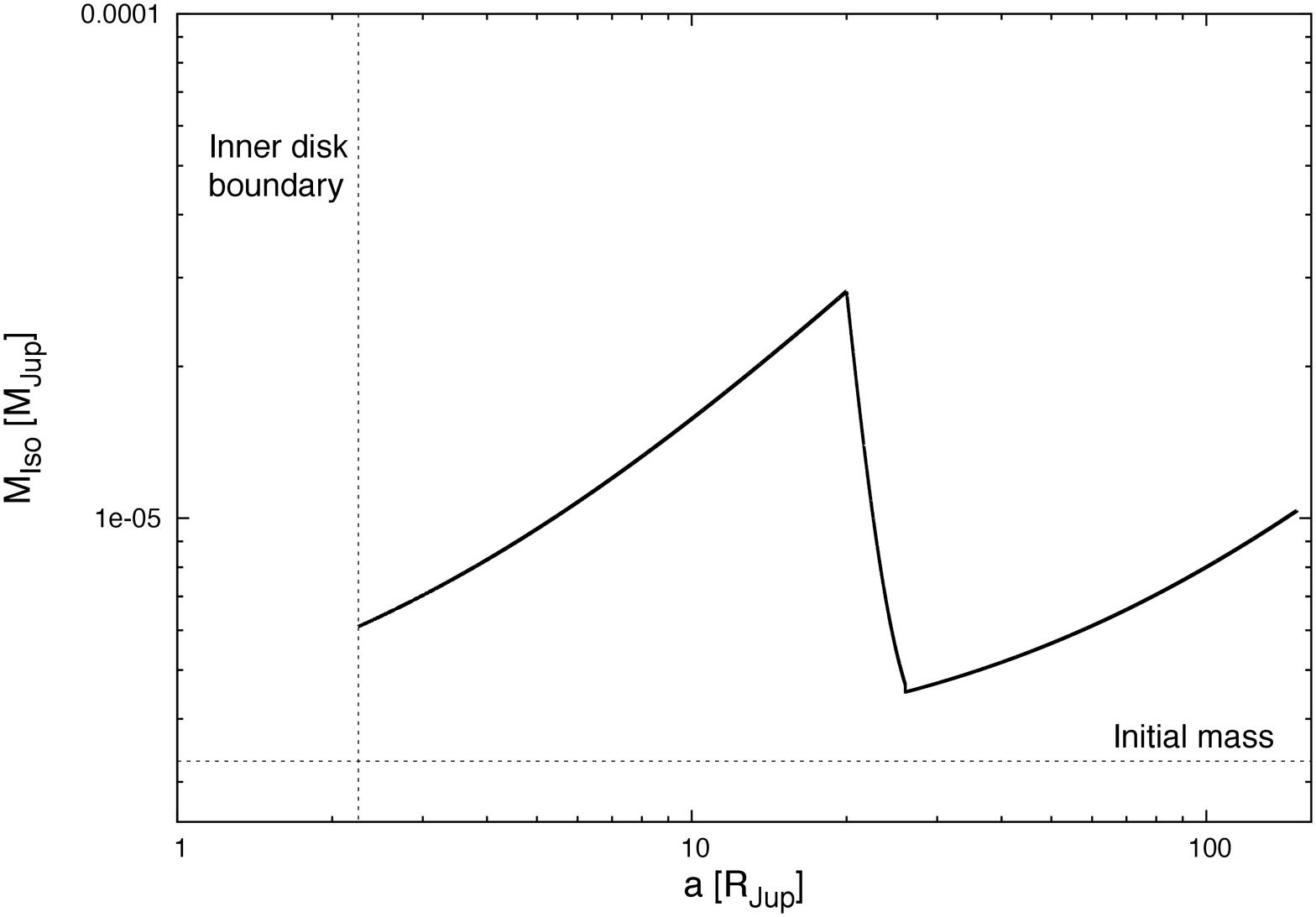}
\caption{Isolation mass as a function of semimajor axis. Dotted lines show inner disk boundary (vertical) and initial satellite-embryos' mass (horizontal) adopted.}
\label{isolation}
\end{center}
\end{figure}

The isolation mass ($M_{iso}$) is the mass reached by a satellite when it accretes all the solids in its feeding zone. The definition of $M_{iso}$ neglects satellite migration and capture of passing planetesimals, but it is a useful concept to interpret the results even in cases when migration is incorporated (see section \ref{results}). 
The isolation mass is given by
\begin{equation}\label{m_iso}
M_{iso}=2\pi a \Delta a \Sigma_s
\end{equation}
with $\Delta a\simeq10R_{H}$ the feeding zone of the satellite-embryo. Figure \ref{isolation} shows $M_{iso}$ as a function of semi major axis. Embryos located at $\sim15~R_{Jup}$ have the largest amount of mass available to grow (comparable to Europa's mass). 

\subsubsection{Satellites Migration}\label{a-evo}

\paragraph{Type I migration} Satellite-embryos migrate due to the interaction with the gaseous disk. Small satellites that are not able to open a gap migrate with a type I migration time-scale given by equation \ref{typeI} \citep{ta02}. 

\begin{equation}\label{typeI}
\tau_{migI}=\frac{1}{2.7+1.1}\left(\frac{c_s}{a \Omega_k}\right)^2\left(\frac{M_{Jup}}{M_{sat}}\right)\left(\frac{M_{Jup}}{\Sigma_g a^2}\right)\Omega^{-1}
\end{equation}
with $c_s$ is the sound speed. Satellite-embryos start migrating when the condition given by equation \ref{condition-migI} is satisfied \citep{sa10,il08}:
\begin{equation}\label{condition-migI}
\tau_{acc} > \tau_{migI} 
\end{equation}
Type I migration time-scale for a 1000 km body located at Europa's semimajor axis is $6\times 10^2$ years and for one located at Ganymede's semimajor axis is $8 \times 10^2$ years, much longer than the accretion time-scales. A comparison with equation \ref{accretionrate} indicates that condition \ref{condition-migI} is only satisfied when the satellite's mass reaches $M_{iso}$, therefore when it stops accreting. In addition, a gap is easily opened up in a low viscosity disk such as the ones considered here. Therefore, in most cases, a satellite that reaches the isolation mass already acquired the necessary mass to open up a gap in the orbit (see equation \ref{gap}) and migrate in the type II regime, as will be discussed next. Since accretion is dominant before reaching $M_{iso}$ and migration switches to type II migration regime after reaching $M_{iso}$, there is no phase where type I migration is effective for orbital evolution. For this reason we do not consider the newly proposed formula of type I migration for non-isothermal disks (e.g., \citet{pa11}).

\paragraph{Type II migration} Tidal interaction between the gaseous disk and the embedded protosatellite induces a truncation of the disk and a gap is opened around the protosatellite's orbit. In a viscous evolving disk, the gas continuously diffuses into the protosatellite, refilling its feeding zone. A gap is opened in the disk when the mass of the protosatellite is big enough to allow that the timescale for the gap opening due to tidal torques is shorter than the timescale on which the viscosity can refill the gap \citep{lp,il04}:

\begin{equation}\label{gap}
\begin{split}
M_{gap}=\frac{40~\nu}{a^2\Omega_k}M_{Jup} 
\sim 40 \alpha \left(\frac{h}{a}\right)^2 M_{Jup} \\
\sim 4 \times 10^{-6} \left(\frac{\alpha}{10^{-5}}\right) \left(\frac{h/a}{0.1}\right)^2 M_{Jup}, 
\end{split}
\end{equation}
where $h$ is the scale height, $\alpha$ is the parameter that characterizes the viscosity, and we used $\nu=\alpha h^2\Omega_k$ \citep{ss}. In a disk in which MRI operates, $\alpha\sim 10^{-3}-10^{-2}$. In this paper, we assume that MRI does not work in circumplanetary disks, and correspondingly, we consider a value between $10^{-6}$ and $10^{-4}$  \citep{mea03a}. Because of the small $\alpha$, satellites open up a gap at relatively small masses, switching to
type II regime before type I migration becomes dominated over growth. \citep{li96,tr02,ar02}. 

Type II migration has two regimes: satellite or disk dominated \citep[e.g.,][]{ha13}. Initially, when the protosatellite's mass is lower than the mass of the disk within the protosatellite's radius, type II migration is in the disk-dominated regime with the same time-scale as the local viscous diffusion time-scale, which is approximated by  \citep{ha13} 
\begin{equation}\label{disk-dominated}
\tau_{migII,d} \simeq \frac{M_{disk}(a)}{\dot{M}_{vis}}
\simeq\left(\frac{a}{R_{disk}}\right) \tau_{disk},
\end{equation}
where $M_{disk}(a)$ is the mass of the disk within the protosatellite location, $R_{disk}$ is a characteristic disk size, $\dot{M}_{vis}$ is the disk mass accretion rate and $\tau_{disk}$ is the gaseous disk global dissipation time-scale. If we assume $R_{disk} \sim 100~R_{Jup}$ and $\tau_{disk}=10^5$ years, the time-scale for type II disk-dominated migration for a satellite like Europa and Ganymede is $\sim 10^4$ years. 

When the protosatellite mass is larger than the mass of the disk within the protosatellite's radius, disk accretion from outer regions pushes the satellite rather than an inner disk and the type II migration switches to the satellite-dominated regime, with a time-scale given by \citep{ha13}:

\begin{equation}\label{eq-migII}
\tau_{migII,s}\simeq sign(a_s-R_m) M_s/\dot{M}_{vis}
\end{equation}
where the subindex "s" means that the quantity is evaluated for the satellite and the subindex "m" is related to the $R_m$, the radius where the type II migration changes direction in the protosatellite disk.  The value of $R_m$ depends on the gaseous disk distribution \citep{il04}, we assume a value given by $R_m\simeq 23~e^{2t/5\tau_{disk}}~R_{jup}$.

We adopt these simplified formulas for type II migration regime, but there are many uncertainties regarding this complex process. Three-dimensional hydrodynamic calculations of the disk torque exerted on a planet embedded on a gaseous disk, showed that the circumplanetary material can induce very strong torques on the planet, slowing down their inward drifting motion \citep{da02,da03}. On the other hand, nonlinearities of the flow around the planet, might also yield to a positive excess of the corotation torque leading to a slowing down or reversal of the migration \citep{ma06}. In order to take into account these uncertainties, we introduce a factor $1/C_{migII}$ in equations \ref{disk-dominated} and \ref{eq-migII}, ($C_{migII} \le 1$) that acts slowing down type II migration. We perform simulations with $C_{migII}=1,0.1$ and $0.01$ (sections \ref{sa-mII} and \ref{section-migII}).  

Due to the presence of a disk cavity, satellites stops migrating when they reach the inner disk boundary. Nevertheless, a satellite located in the inner boundary may be pushed inward by another migrating satellite in a 2:1 resonance. Because the disk inner edge is likely to correspond to the corotation radius, the satellite pushed inward should tidally decay to the planet. In N-body simulations by \citet{og12}, satellite's migration is halted at the inner disk edge without being pushed inside of the cavity and subsequently migrating satellites are trapped in outer resonances, because a strong torque is exerted on the innermost satellite ("eccentricity trap") to prevent the satellite system from falling onto the planet \citep{og10}. However, the strong edge torque is caused by dynamical friction and it only works in the case without gap opening. In the low-viscosity disk that we are considering, the gap is opened easily and therefore the "eccentricity trap" does not occur.

Satellites also stop migrating when the disk is severely depleted of gas (see section \ref{gas_evolution}).
 
\subsubsection{Resonance Trapping}

There are several satellite-embryos growing and migrating simultaneously in the disk. Some of them might enter a mean motion resonance. Since Galilean satellites are in  resonances  (Io-Europa and Europa-Ganymede are in 2:1 mean motion resonance), this effect is highly relevant for Jovian satellites formation and evolution.

We compute the dynamical perturbation between two protosatellite neighbors neglecting the perturbation of other distant objects. When the neighbors are in nearly circular orbits, the expansion of $db$, the difference in their semimajor axes, was calculated by \citet{gt82} and \citet{hn90}. Since the encounters occur at every synodic period, the change in the rate is given by equation \ref{resonance} \citep{il10}. 

\begin{equation}\label{resonance}
\frac{db}{dt} \simeq \frac{\delta b}{T_{Syn}} \simeq 7\left(\frac{b}{r_{H}}\right)^{-4}\left(\frac{r_{H}}{a}\right)^2v_K
\end{equation}
with $r_{H}$ the Hill radius and $v_K$, the Kepler velocity. When $db/dt$ becomes comparable to the difference in migration speed ($db/dt=\Delta v_{mig}$), the interaction between the converging satellites should be compensated by their relative motion. The separation of the satellites in this equilibrium ($b_{trap}$) is maintained at a distance given by equation \ref{eq}. 

\begin{equation}\label{eq}
b_{trap} \simeq 0.29 \left(\frac{M_{sat}}{10^{-4}M_{Jup}}\right)^{1/6}\left(\frac{\Delta v_{mig}}{v_K}\right)^{-1/4}R_H.
%\label{eq:b_trap}
\end{equation}
If type II migration rate is used for $\Delta v_{mig}$ and Europa like satellites  
are assumed, $b_{trap}\sim 11~R_H$.

We include resonance trapping in a simplified way assuming that two neighbors, convergent protosatellites will enter in a low order main motion resonance with a distance between them given by $b_{trap}$. A similar approach was used in \citet{sa10}, a more detail resonance trapping model is out of the scope of our work. 

Note that resonant trapping does not occur for satellitesimals that are small enough to migrate due to gas drag. Migration timescale of satellitesimals drift towards Jupiter is so short that the estimate $b_{trap}$ by eq.~(\ref{eq:b_trap}) is smaller than $R_H$ (see section \ref{nebula-evolution}), therefore they can not be resonantly trapped. We consider resonant trapping only between satellites migrating with type II migration.

\subsection{Gaseous disk evolution}\label{gas_evolution}

\begin{figure}[ht]
\begin{center}
\includegraphics[angle=90,scale=.28]{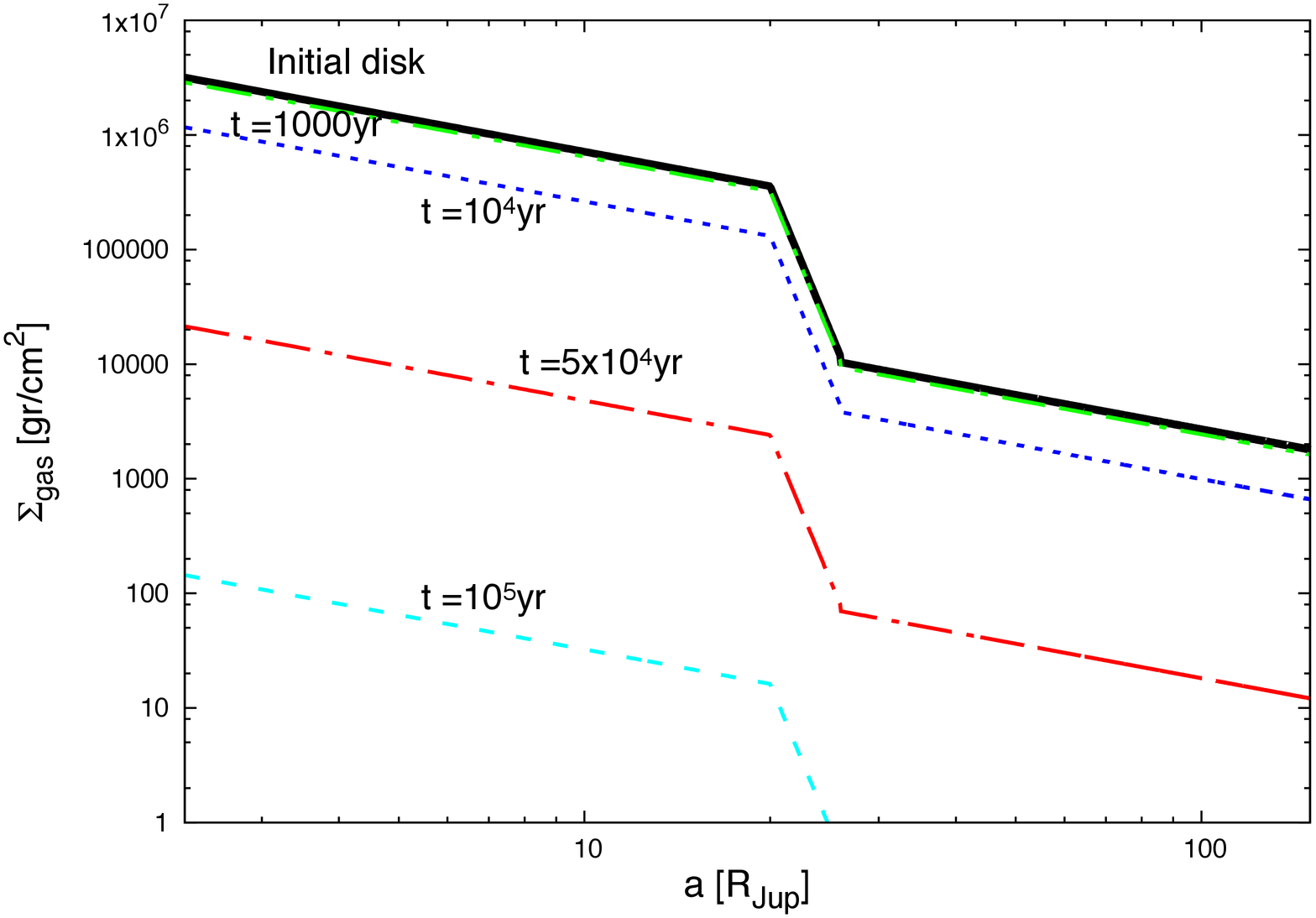}
\caption{Gaseous disk surface density vs. semimajor axis as it evolves with time. Different lines show $\Sigma_{gas}(a)$ at different times: initial (black solid), 1000 (green dashed), $10^4$ (blue dotted), $5\times10^4$ (red dotted and dashed) and $10^5$ years (cyan dashed line).}
\label{gas-disk-ev}
\end{center}
\end{figure}

Due to the viscous diffusion, the gaseous disk component of the Jovian protoplanetary disk decreases with time.
The global disk viscous diffusion time-scale is given by
\begin{equation}
\tau_{disk} \sim \frac{R_{disk}^2}{\alpha h^2 \Omega_K}
                     \sim 3 \times 10^5 \frac{R_{disk}}{100R_{Jup}} 
                                  \left(\frac{\alpha}{10^{-5}}\right)^{-1} {\rm yr}.
\end{equation}
An study by \citet{fu14} showed the absence of MRI-active regions 
in the circumplanetary disks, suggesting a value of $\alpha$ much smaller than $10^{-3}-10^{-2}$.
Thus, we adopt $\tau_{disk}$ between $10^{4}$ to $10^7$yr in our simulations. This time-scale is longer than the satellite formation time-scale (equation \ref{accretionrate}), but it is important to consider the orbital evolution of the satellites. We adopt an approximate prescription for the global gaseous disk dissipation in our calculations given by \citep{il04}:

\begin{equation}\label{gas-dis}
\Sigma_g\simeq \Sigma_{g,0}~e^{-\frac{t}{\tau_{disk}}}
\end{equation}
 with $t$ the time and $\Sigma_{g,0}$ the initial gas surface density. The evolution of the gaseous disk with time is shown in figure \ref{gas-disk-ev}. In the figure we adopted $\tau_{disk}=10^4$ years and show the gas surface density at $1000$, $10^4$, $5\times10^4$ and $10^5$ years with green dashed, blue dotted, red dotted and dashed and cyan dashed lines, respectively.  

\subsection{Satellitesimal disk evolution}\label{nebula-evolution}

\begin{figure*}
  \begin{center}
\subfigure[]{\label{5}\includegraphics[angle=90,width=.33\textwidth]{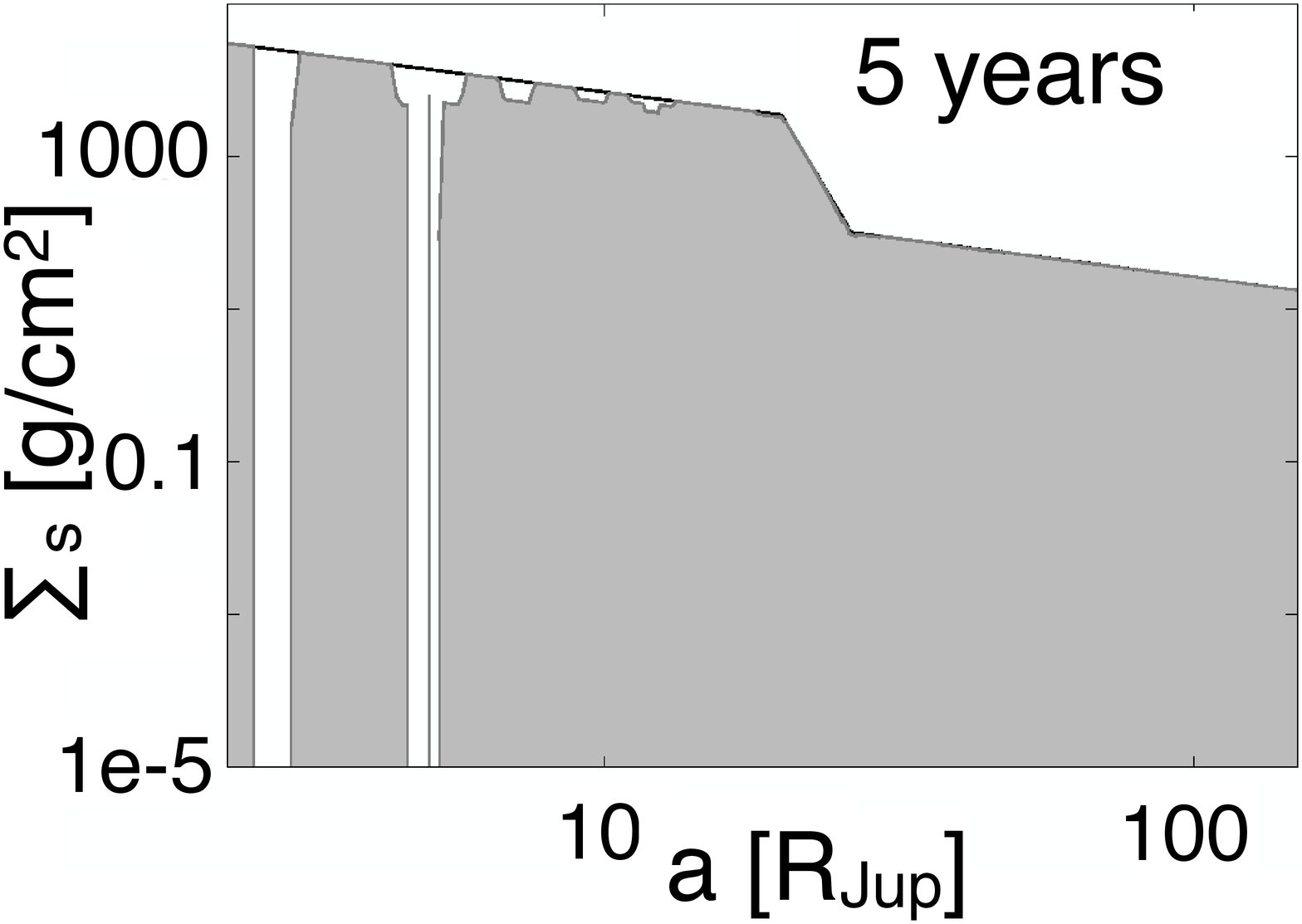}}\subfigure[]{\label{10}\includegraphics[angle=90,width=.33\textwidth]{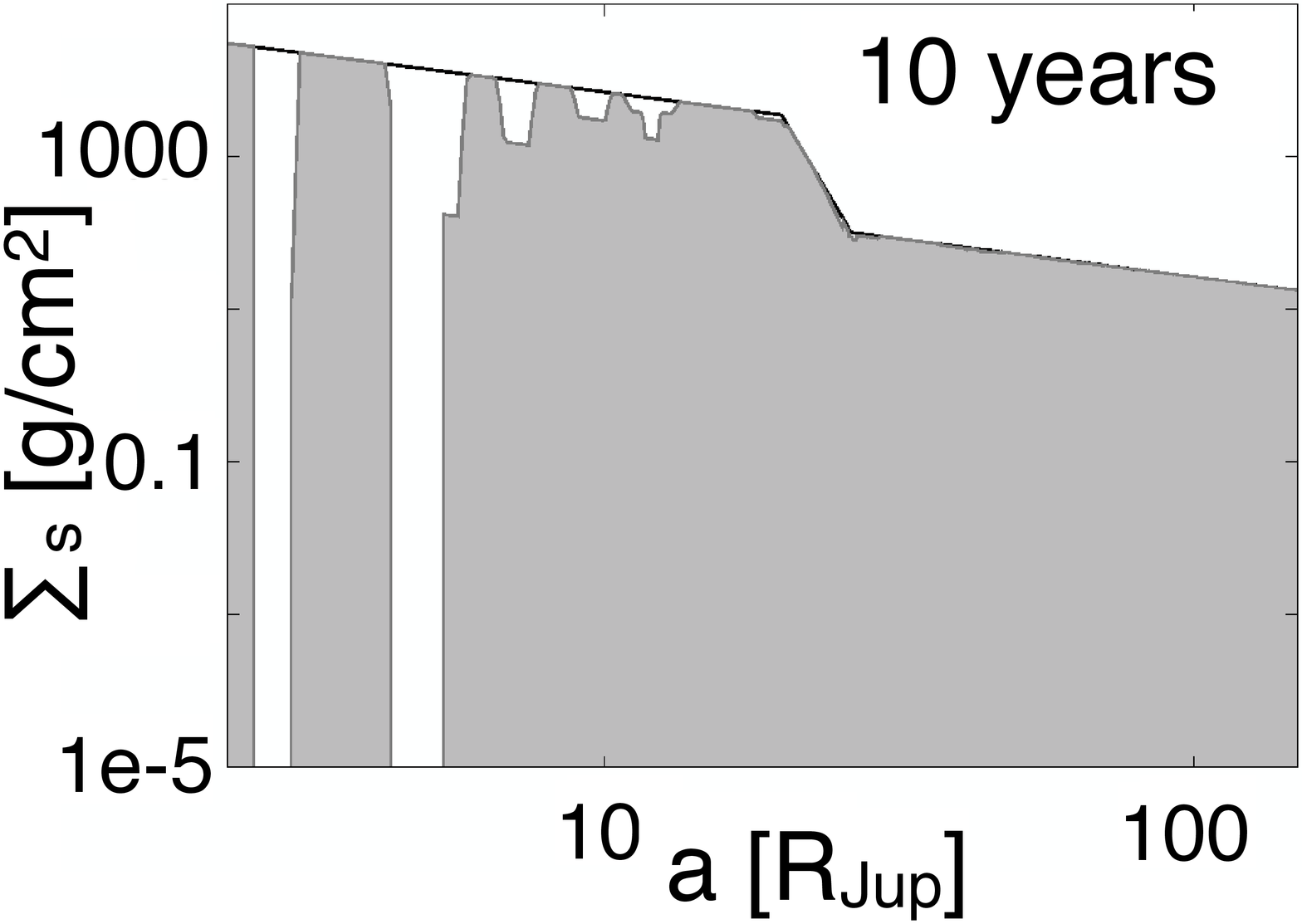}}\subfigure[]{\label{100}\includegraphics[angle=90,width=.33\textwidth]{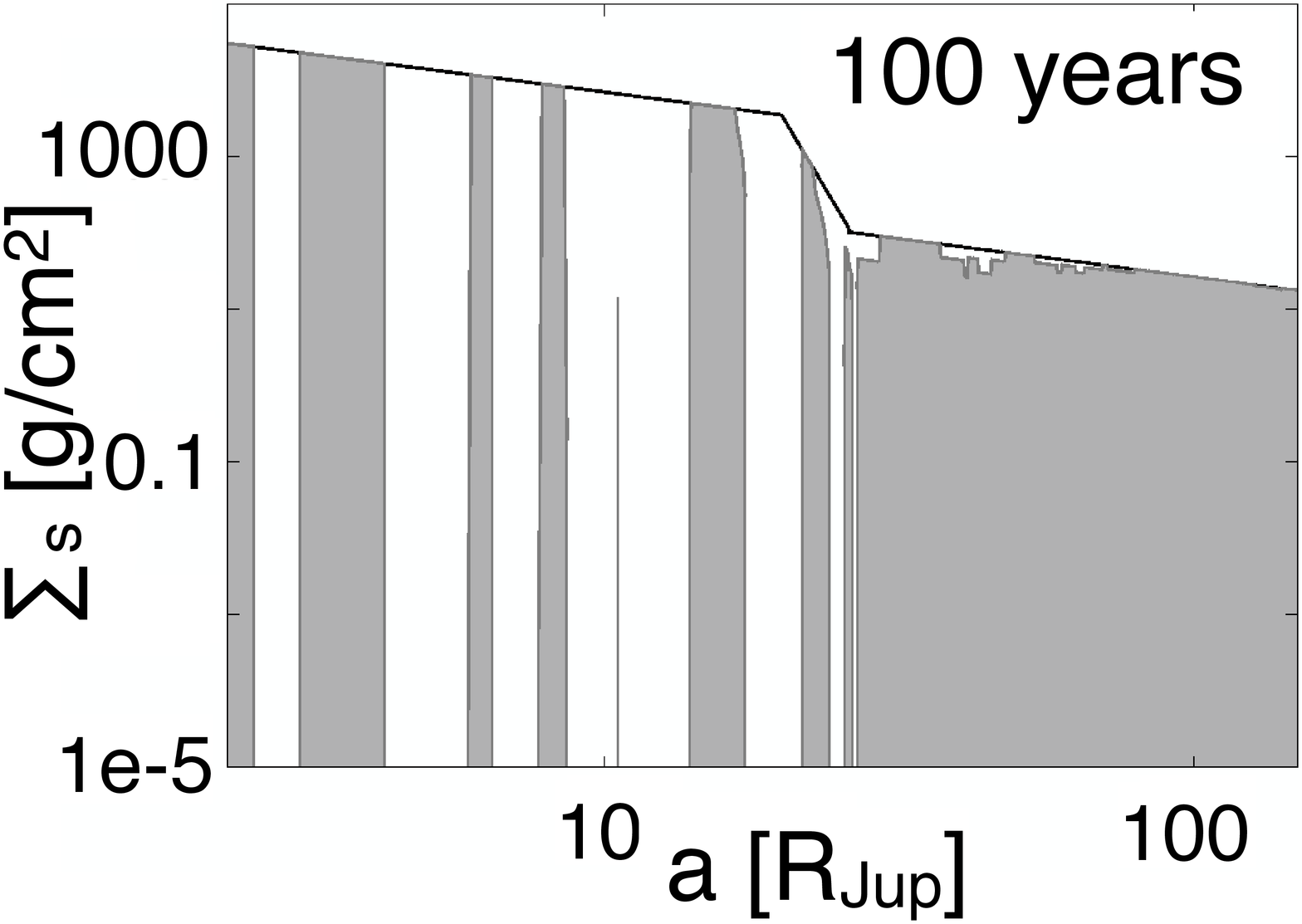}}
\subfigure[]{\label{500}\includegraphics[angle=90,width=.33\textwidth]{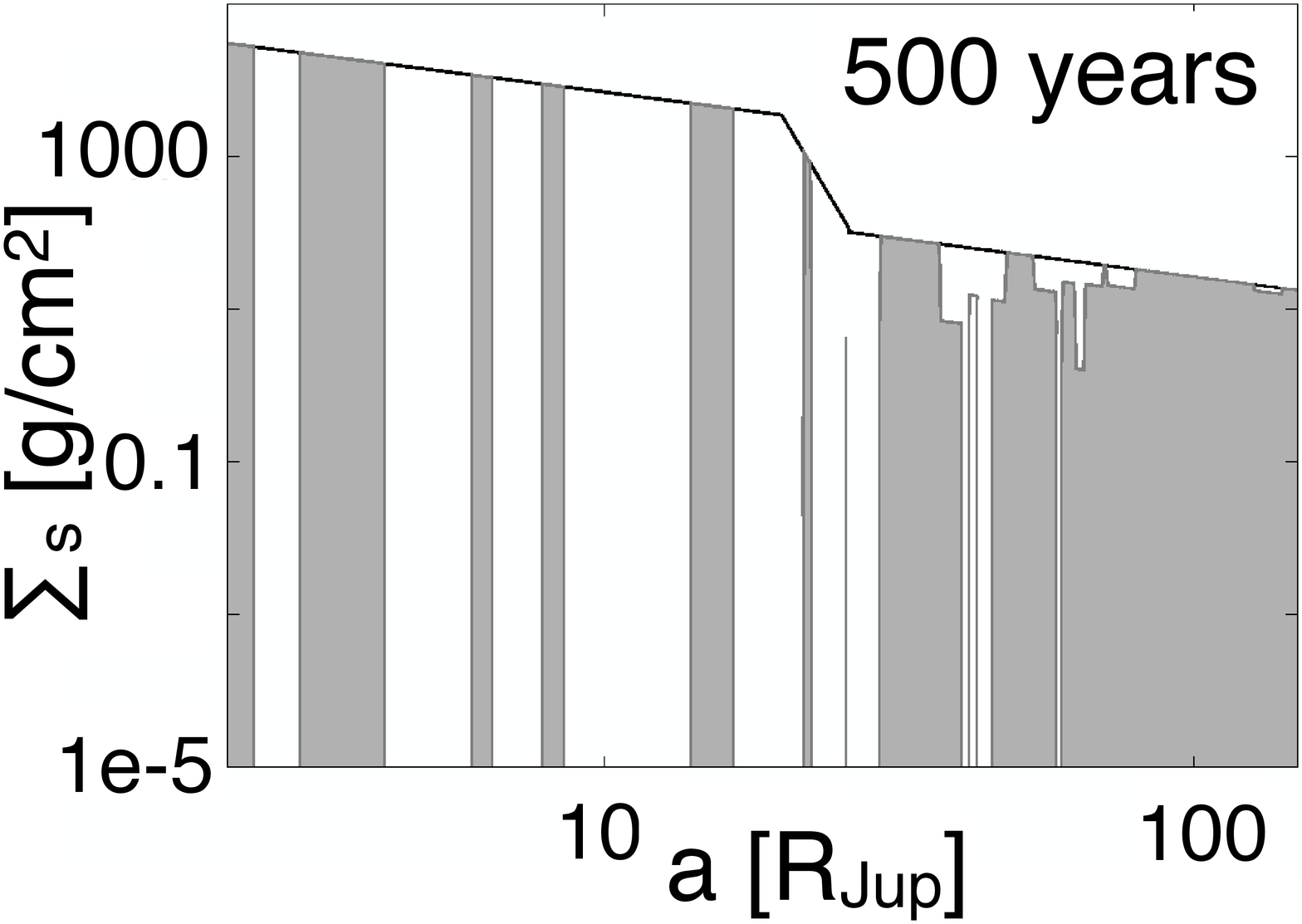}}\subfigure[]{\label{1000}\includegraphics[angle=90,width=.33\textwidth]{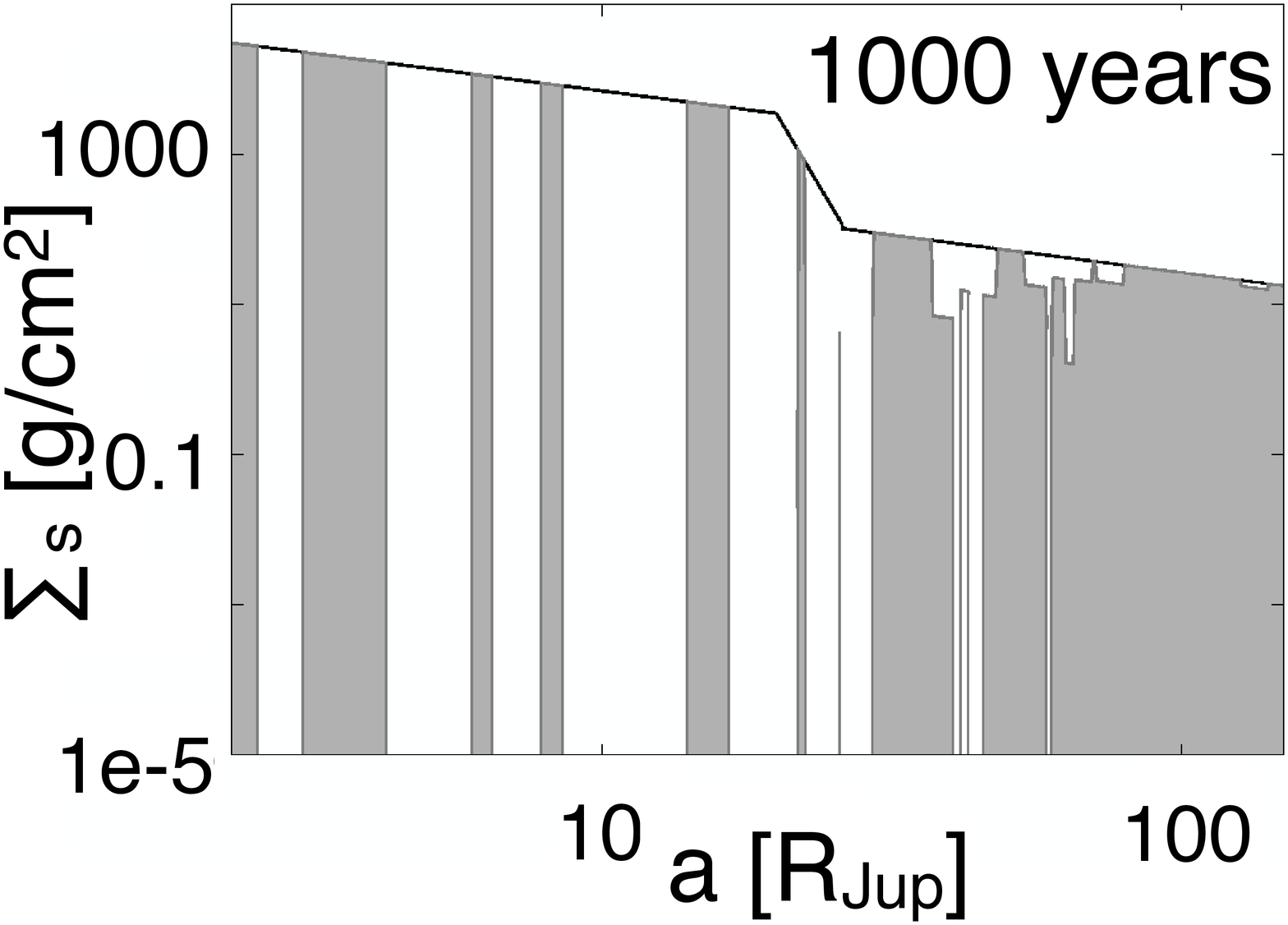}}\subfigure[]{\label{5000}\includegraphics[angle=90,width=.33\textwidth]{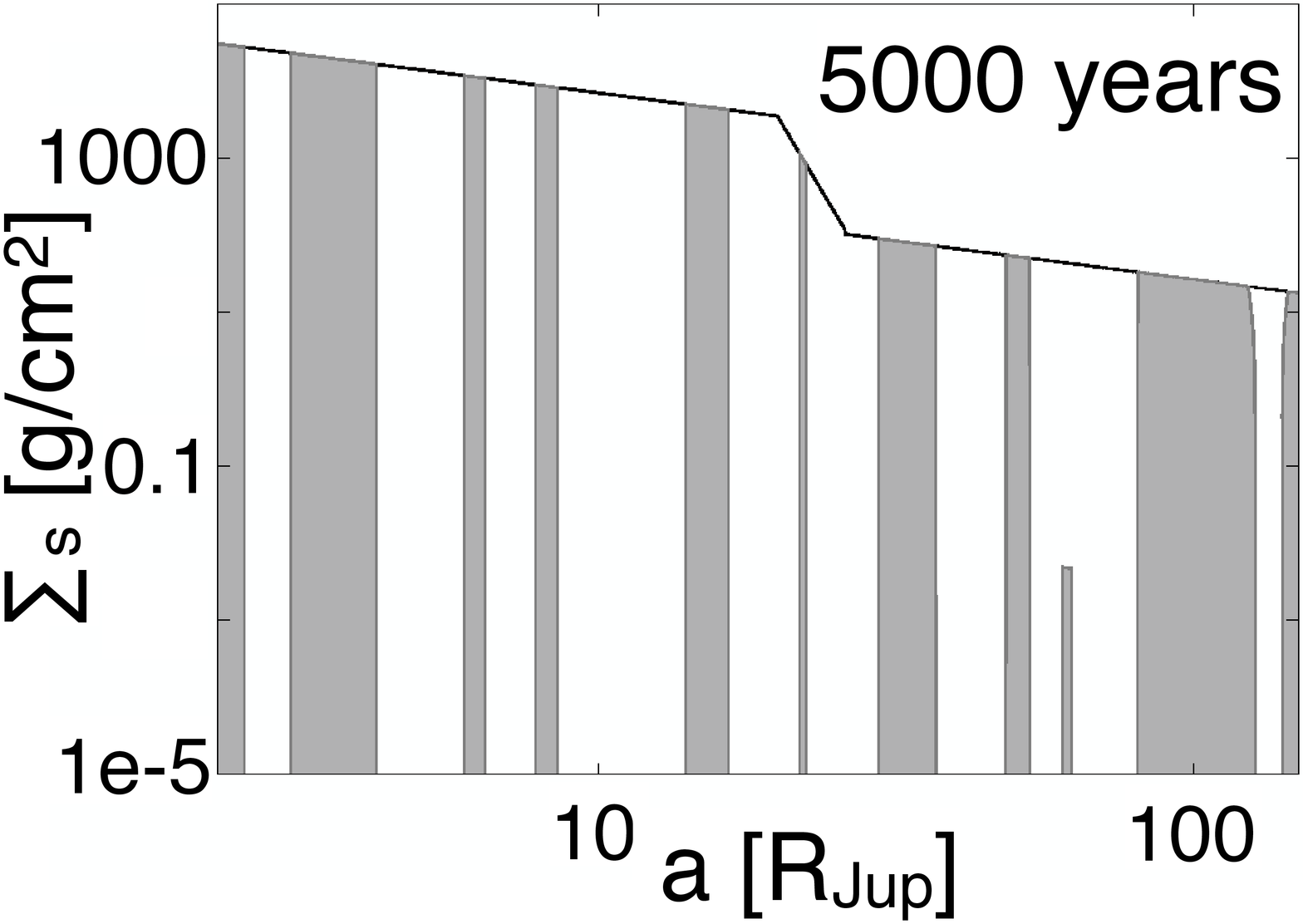}}
 \end{center}
  \caption{Solids surface density evolution as a function of semimajor axis at different times: 5 (\ref{5}), 10 (\ref{10}), 100 (\ref{100}), 500 (\ref{500}), 1000 (\ref{1000}) and 5000 (\ref{5000}) years. The depletion is due to the accretion of in-situ satellite-embryos.}
  \label{disc-evolution-growth}
\end{figure*}

Satellites stop growing when the Jovian nebula has been severely depleted of solids. There are two processes that deplete the solid disk: depletion of small satellitesimals due to protosatellites accretion and drift of satellitesimals towards Jupiter due to gas drag effect.

\paragraph{Depletion due to protosatellites accretion} 
Satellitesimals are accreted into the protosatellites, depleting protosatellites' feeding zones. The surface density after protosatellite accretion is given by equation \ref{solids1}:

\begin{equation}\label{solids1}
\Sigma_s=\Sigma_{s,0}-\frac{M_{sat}}{2\pi a~10 R_H}
\end{equation}
with $\Sigma_{s,0}$ the initial surface density. 

Figure \ref{disc-evolution-growth} shows the solids surface density evolution as a function of semimajor axis. The change in the solid disk at different times is due to the accretion into the protosatellites. In this result, we neglected migrations of both satellite-embryos and satellitesimals.
Embryos in the inner region accrete all the material in their feeding zones in less than $\sim$100 years, while the embryos located in the outer disk grow more slowly. 

\paragraph{Satellitesimals migration due to gas drag} Disk gas orbits Jupiter at a lower velocity than a solid body moving on a Keplerian orbit. Therefore an object orbiting at Keplerian speed suffers a gas drag and drifts toward the planet. This effect is important for the small satellitesimals in the disk, before they become large and decouple from the gaseous nebula. 

\begin{figure*}
  \begin{center}
\subfigure[]{\label{drag-1km}\includegraphics[angle=90,width=.42\textwidth]{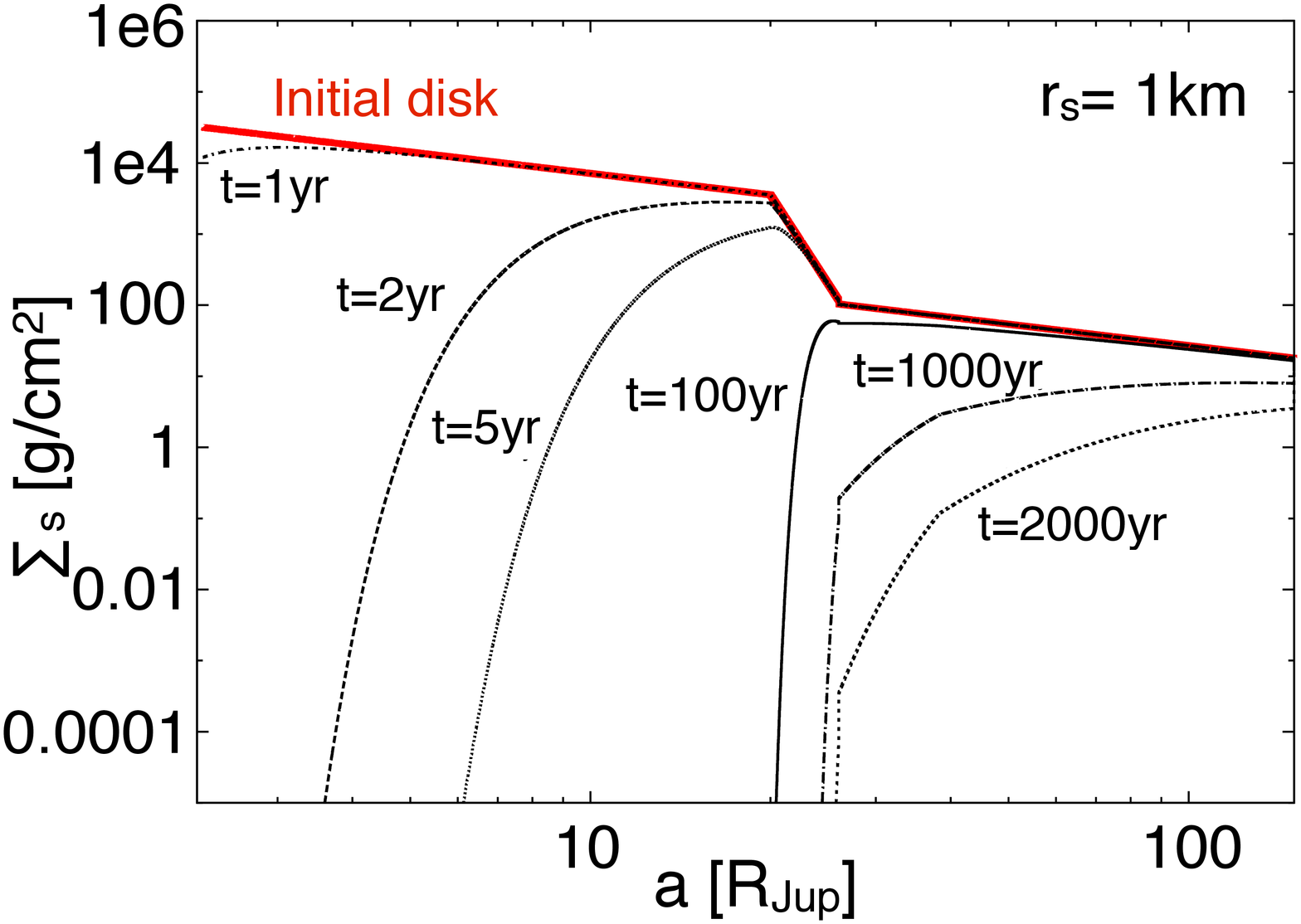}}\subfigure[]{\label{drag-10km}\includegraphics[angle=90,width=.42\textwidth]{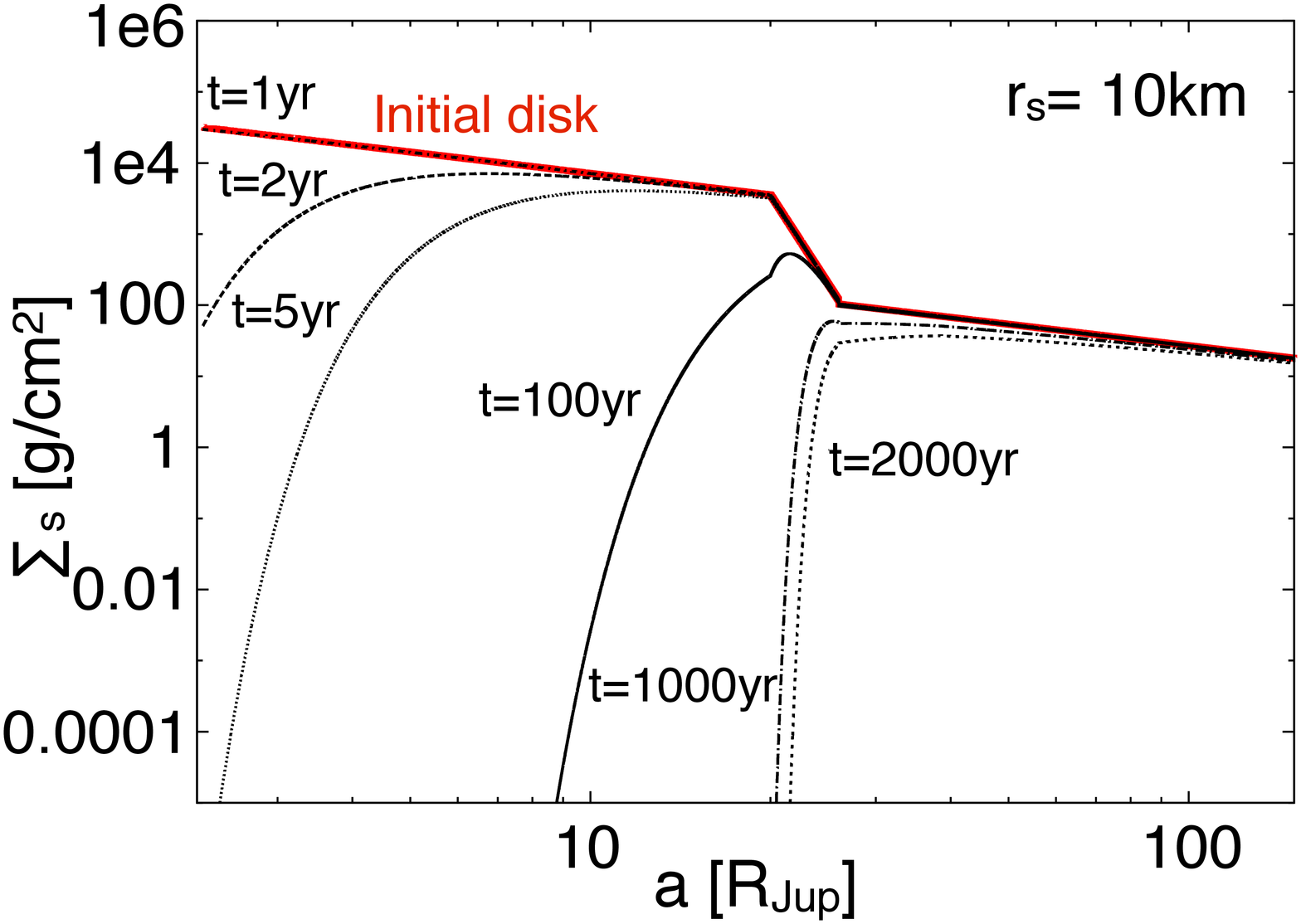}}
\subfigure[]{\label{drag-20km}\includegraphics[angle=90,width=.42\textwidth]{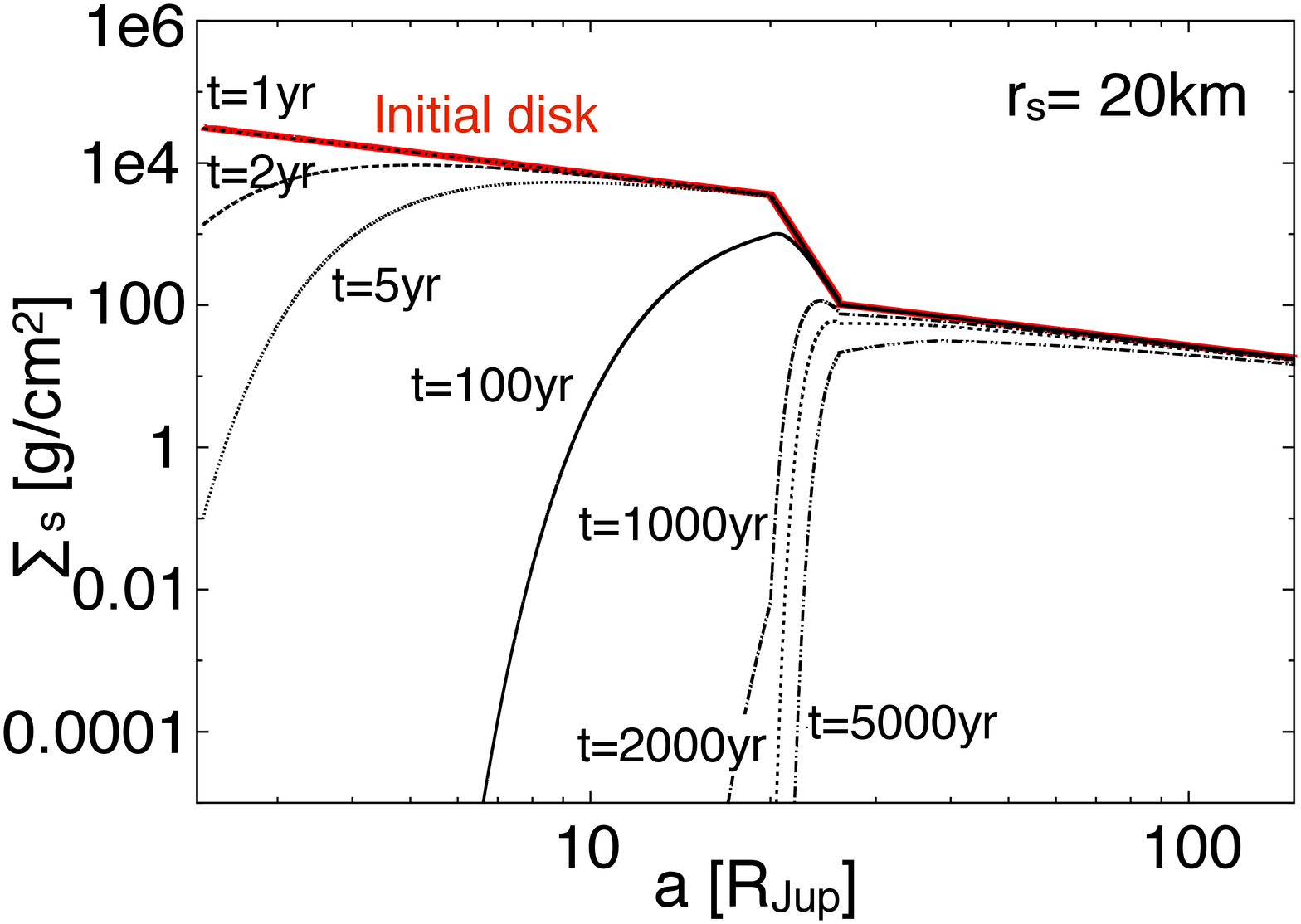}}\subfigure[]{\label{drag-30km}\includegraphics[angle=90,width=.42\textwidth]{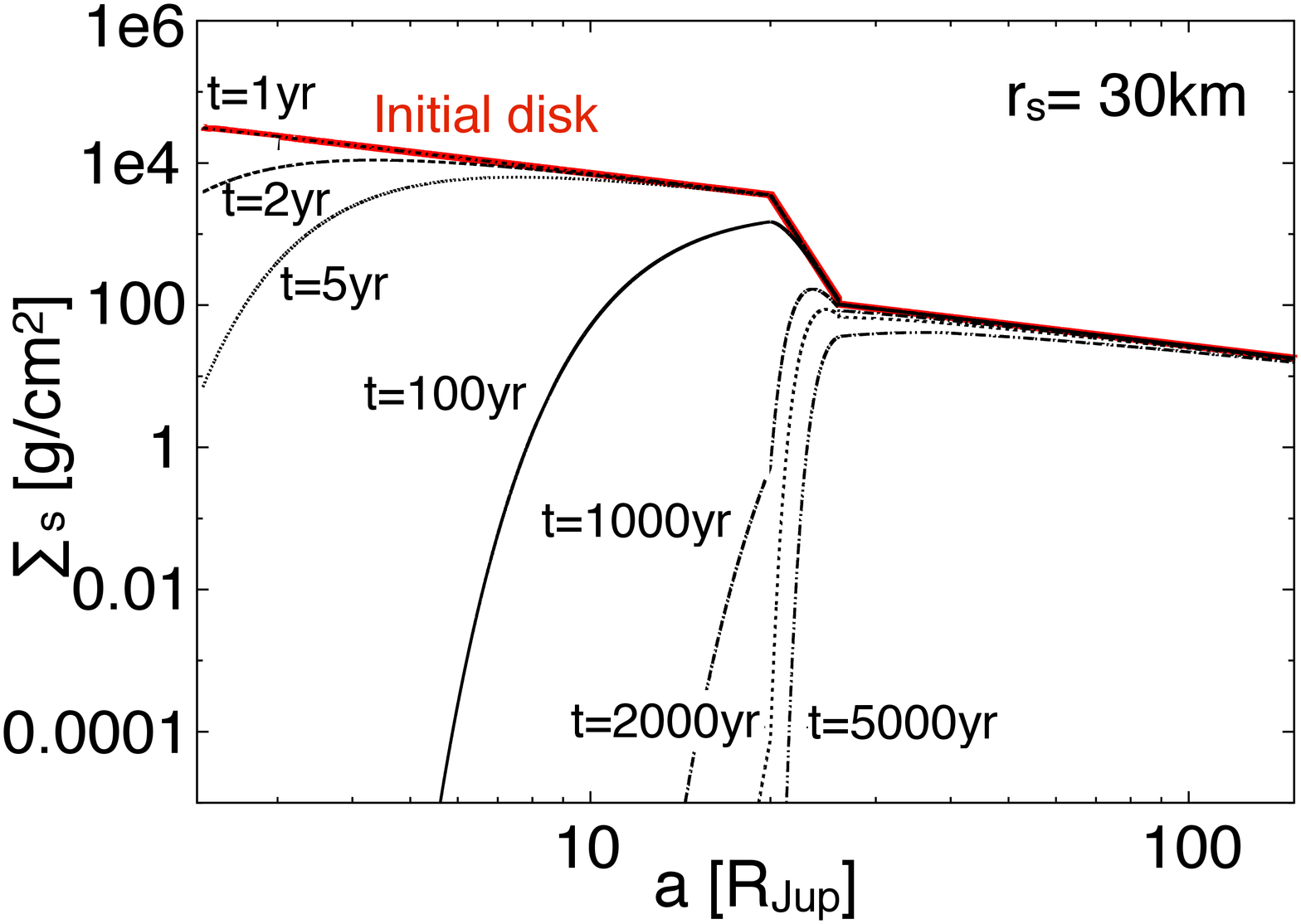}}
 \end{center}
  \caption{Evolution of solid disk due to gas drag effect, when gas to dust ratio is 100. Different Figures show the results when adopting different satellitesimals size: 1, 10, 20 and 30km (Figures \ref{drag-1km},\ref{drag-10km},\ref{drag-20km} and \ref{drag-30km}, respectively). Initial solid disk is shown (red solid). Evolution of the disk at different times are shown (1,2,5, 100, 1000, 2000 and 5000 years).}
  \label{disc-evolution-gas100}
\end{figure*}

The satellitesimal migration time-scale ($\tau_{gas}$) due to gas drag effect is given by equation \ref{drag} \citep{mea03a}.
\begin{equation}\label{drag}
\tau_{gas}=22\left(\frac{\rho_s}{1\frac{g}{cm^{3}}}\right)\left(\frac{r_{s}}{1~km}\right)\left(\frac{T}{130~K}\right)^{-\frac{3}{2}}\left(\frac{\Sigma_g}{10^5 \frac{g}{cm^{2}}}\right)^{-1}~{\rm yr}
\end{equation}
where $\rho_s$ and $r_s$ are the typical satellitesimals density and radius. Equation \ref{drag} shows that a satellitesimal of $1$ km and density equal to $1~g/cm^2$ initially located at $6R_{Jup}$ reaches the inner disk boundary only in $\sim 0.2$ years, while a satellitesimal initially located in the outer edge of the disk takes $\sim 1200$ years to fall into Jupiter. 

Applying continuity to the solids disk we found that satellitesimals migration change the solids surface density of the Jovian nebula at a rate given by equation \ref{solids2} \citep{th03,mb09}

\begin{equation}\label{solids2}
\frac{\partial \Sigma_s}{\partial t}= -\frac{1}{a}\frac{\partial}{\partial a}\bigg(a \Sigma_s \frac{da}{dt}\bigg)
\end{equation}

Figure \ref{disc-evolution-gas100} shows the solid disk's evolution at different times due to this effect: 1, 2, 5, 100, 1000, 2000 and 5000 years. Red solid line shows the initial solid's disk, black lines show the evolution of the solid's disk at different times. Since drift velocity depends on satellitesimals' size (equation \ref{drag}), we explore the evolution of the disk adopting different radius for the satellitesimals: 1km (Fig. \ref{drag-1km}), 10 km (Fig. \ref{drag-10km}), 20 km (Fig. \ref{drag-20km}), and 30 km (Fig. \ref{drag-30km}). 

As already noted, the accretion time-scale of an Europa/Ganymede-like satellite is 10--15 yr for full retention of satellitesimals. When considering a solid disk with satellitesimals of $< 10$ km in size, the solid disk is depleted earlier than the accretion of the satellites, meaning that satellites-embryos 
do not grow near Europa/Ganymede's semi-major axes (see section \ref{sa-md}).

Satellitesimals' migration is halted near the inner disk edge. As a consequence of this pile up of solid material, a new satellite forms rapidly. This satellite starts type II migration, reaches the inner disk limit and will ultimately decay towards Jupiter, as explained in section \ref{a-evo}. We include these events in a simplified way by removing satellitesimals that reach the inner disk edge.

\section{Results}\label{results}

\subsection{Forming one satellite system under different conditions}
In this section we explore the formation of one satellite system, changing the conditions for its formation, analyzing the implementation of each effect on the results.

\subsubsection{Parameters adopted in the simulations of one satellite system} \label{parameters}
Each simulation starts with 20 satellite-embryos of 1000 km in size located randomly in the disk. Satellite-embryos grow by the accretion the solids in their feeding zone and when they start to migrate (in the cases when we consider migration), this feeding zone moves with them, providing new material to accrete. 

The inward migration of satellite-embryos causes that some of them will move inside the inner disk limit and fall into Jupiter. Nevertheless, if there are solids in the disk, a new generation of satellites can grow from the residual satellitesimals. In order to simulate this process in our calculations and allow the formation of new generations of satellites in the disk, once a satellite moves inside the inner disk limit a new seed is located at the initial location of the satellite that is lost. To represent the mass growth of the residual satellitesimals population, we follow the procedure described by \citet{il10} and adopt for the new seed a mass that is 100 times lower than the one of its predecessor, if enough solid mass remains to create the new seed. As will be shown later on (see section \ref{sa-mII}) a different prescription for accounting the potential growth of subsequent generations does not affect the final results, because in practice, the fast satellitesimals migration ensures that there will be no residual satellitesimals to form new generations of satellites.

In these simulations the solid disk is formed by small satellitesimals of 30 km. For computational reasons, the dissipation time-scale for the gaseous disk is always taken as $10^4$ years \citep{mea03a} and the simulations stop when this time is reached. To represent a low-viscosity disk, the $\alpha$ parameter is adopted as $10^{-5}$ \citep{mea03a}. Different values for the satellitesimals size, dissipation time-scale and $\alpha$ parameter will be explored in the population synthesis calculations (see section \ref{pop}). In all the simulations we consider the depletion of the solids disk due to the accretion of the satellite-embryos. 

\subsubsection{Satellitesimal Accretion only}\label{sa}

\begin{figure}
\begin{center}
\includegraphics[angle=90,scale=.28]{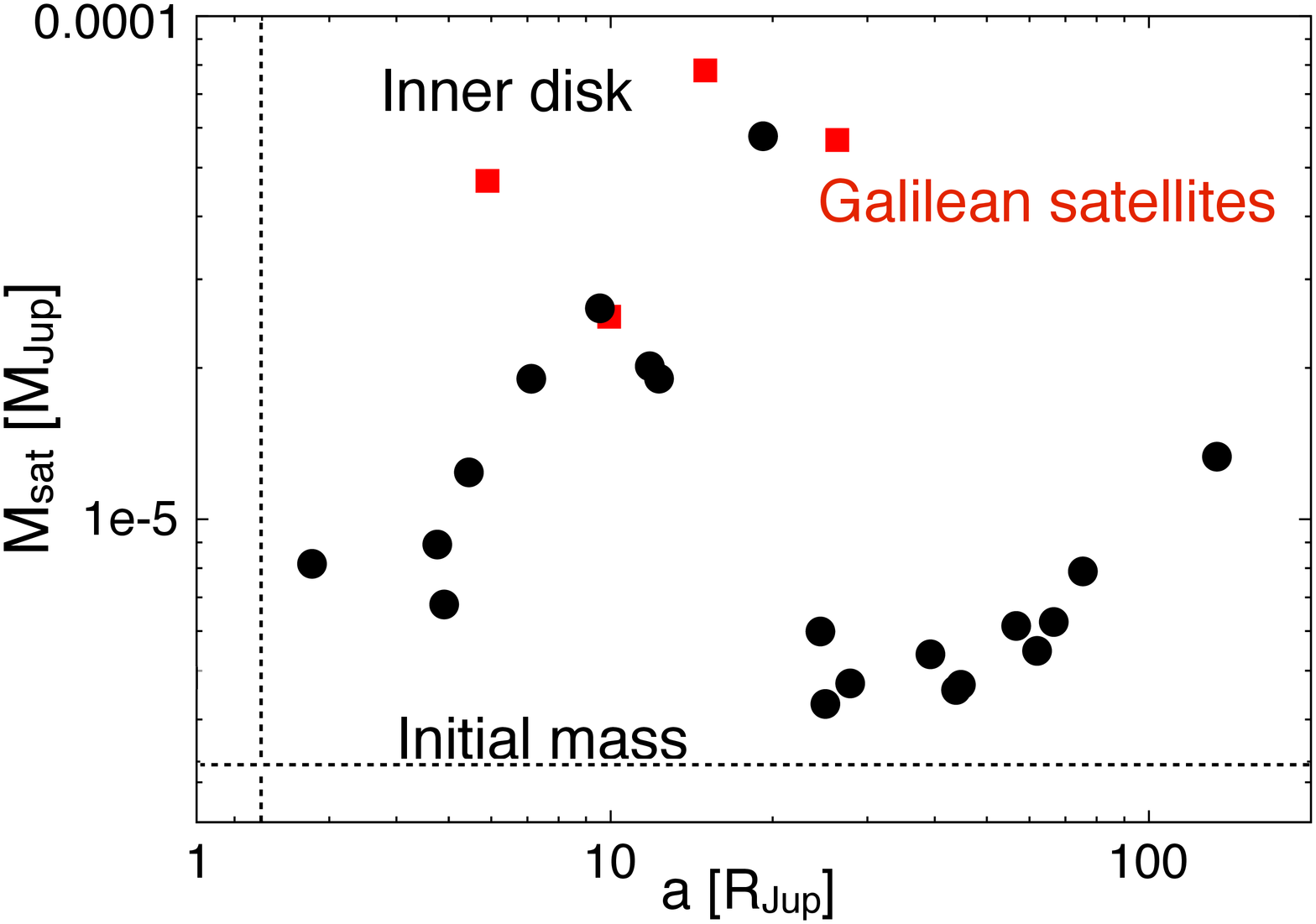}
\caption{Masses versus semi major axis of satellites formed due to accretion in situ. Galilean Satellites are shown (red squares). Doted lines show the inner disk boundary at 2.5 $R_{Jup}$ and initial mass of the satellite embryos.}
\label{manot}
\end{center}
\end{figure}

Figure \ref{manot} shows a resulting satellite system when considering accretion only. No migration (of the satellite-embryos or satellitesimals) is included in this simulation. The largest satellites are located at $\sim15 - 20~R_{Jup}$, in agreement with the calculations of the isolation mass (see Figure \ref{isolation}). 

\begin{figure*}
  \begin{center}
\subfigure[]{\label{growth}\includegraphics[angle=90,width=.42\textwidth]{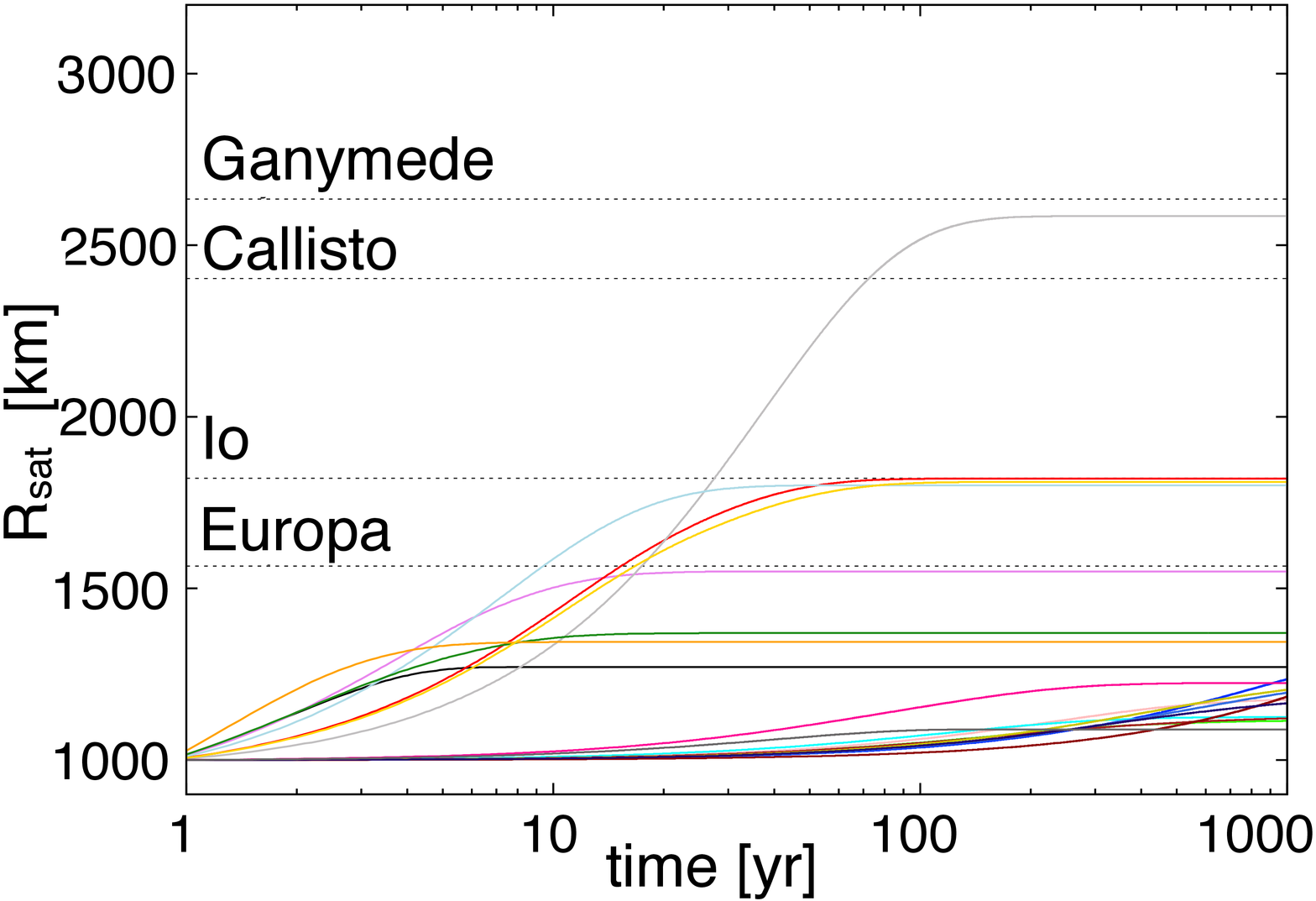}}\subfigure[]{\label{growth-a}\includegraphics[angle=90,width=.42\textwidth]{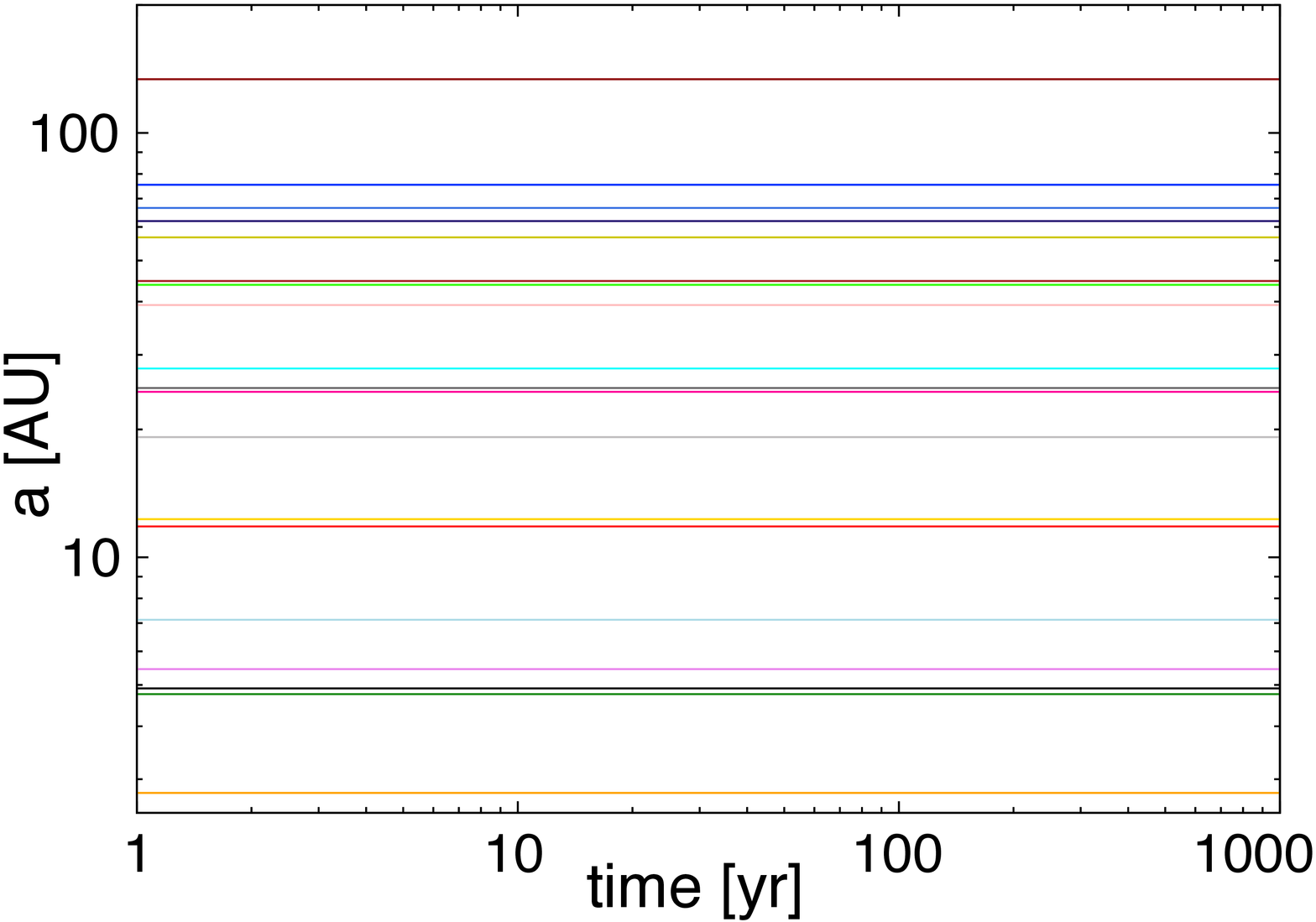}}
 \end{center}
  \caption{Radius vs. time (Figure \ref{growth}) and semimajor axis vs time (Figure \ref{growth-a}) of satellites formed due to accretion in situ. Results up to 1000 years are shown. Galilean satellites' radii are shown for comparison (dotted lines).}
  \label{growth-not}
\end{figure*}

Figure \ref{growth} shows the evolution of radius up to 1000 years for satellites shown in Figure \ref{manot}. Galilean satellites' radii are shown for comparison. In Figure \ref{growth-a} we show the evolution of semimajor axis for these satellites as a reference (semimajor axis do not change with time because satellites do not migrate in this case). Satellites located in the inner region of the disk grow faster and stop growing when they consume all the solids in their feeding zones. The growth is slower for satellite-embryos located at larger semimajor axis.

\subsubsection{Satellitesimal Accretion  + Gas Drag}\label{sa-md}

\begin{figure}
\begin{center}
\includegraphics[angle=90,scale=.28]{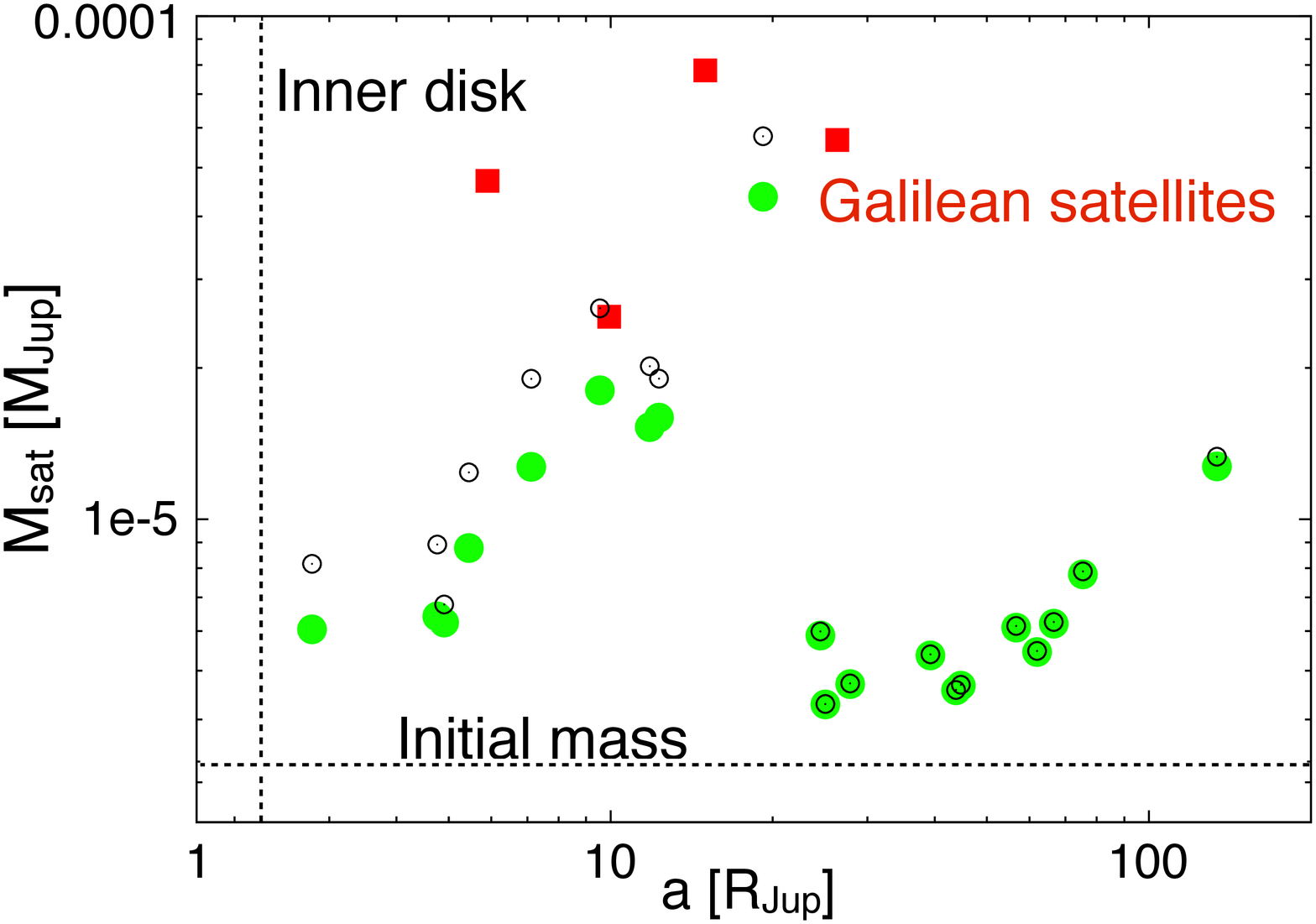}
\caption{Mass vs. semi major axis of satellites formed by accretion in situ, in a disk that evolves due to satellitesimals drag and subsequent drift towards Jupiter. Black open circles are the case of growth without satellitesimals drift and green dots are the case with satellitesimals drift towards Jupiter.}
\label{mamd}
\end{center}
\end{figure}

Figure \ref{mamd} shows the mass vs. semi major axis of the satellite system formed with the same conditions as in section \ref{sa}, but taking into account migration of satellitesimals due to the drag produced by the nebular gas (green dots). The case with no gas drag is plotted for comparison (black open circles). Satellitesimals drift towards the star, depleting the solid disk. Therefore satellites formed when including this effect are smaller. The outer disk takes longer to be depleted, giving satellites more time to grow, implying that satellites located in the outer disk are not so affected. 

\subsubsection{Satellitesimal Accretion  +  migration}\label{sa-mII}

\begin{figure}
\begin{center}
\includegraphics[angle=90,scale=.28]{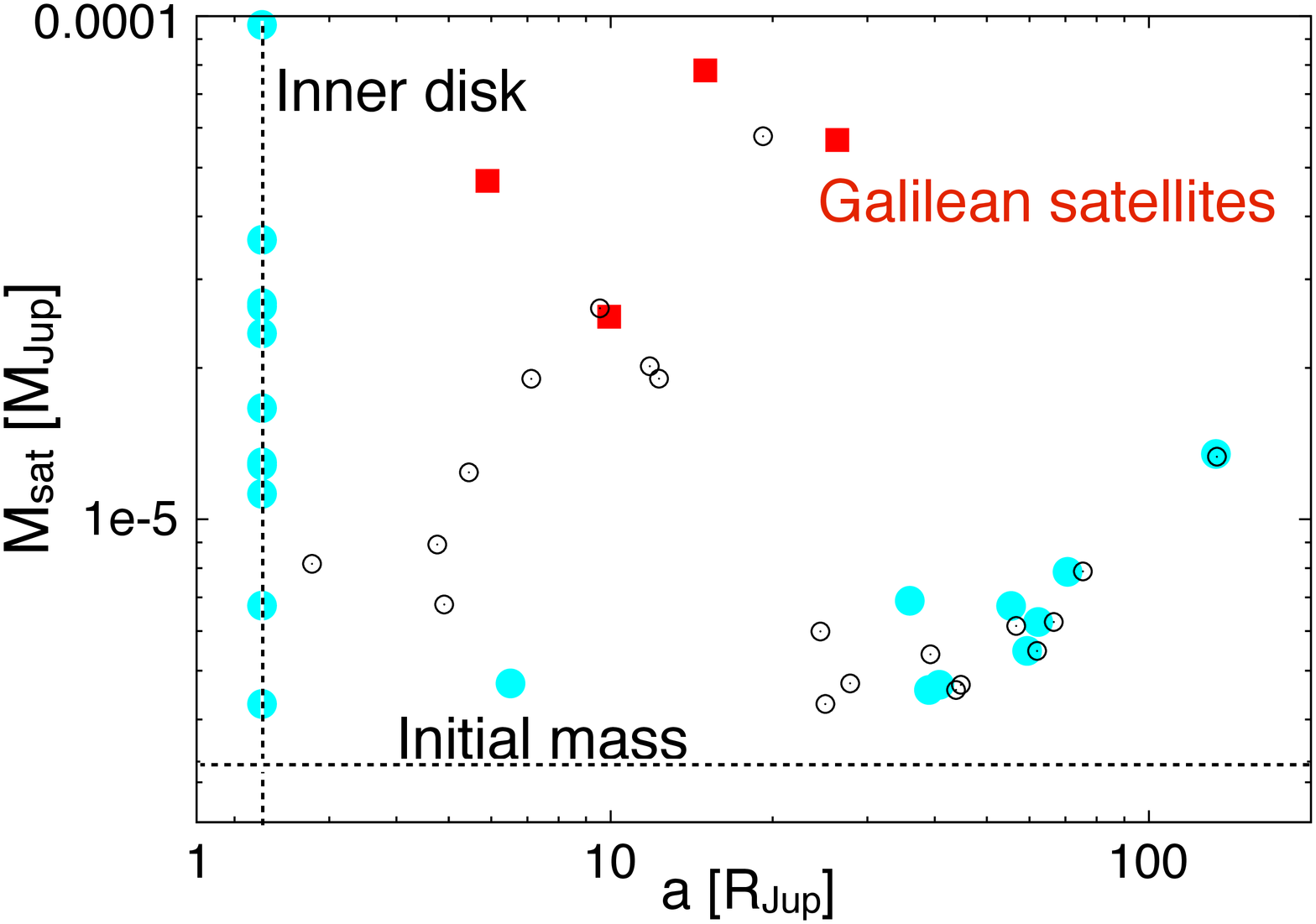}
\caption{Mass vs. semi major axis of satellites formed when considering satellites migration. Black open circles are formed in situ and light blue dots are satellites formed when migration is included.}
\label{mamII}
\end{center}
\end{figure}

\begin{figure*}
  \begin{center}
\subfigure[]{\label{r-t-migII}\includegraphics[angle=90,width=.42\textwidth]{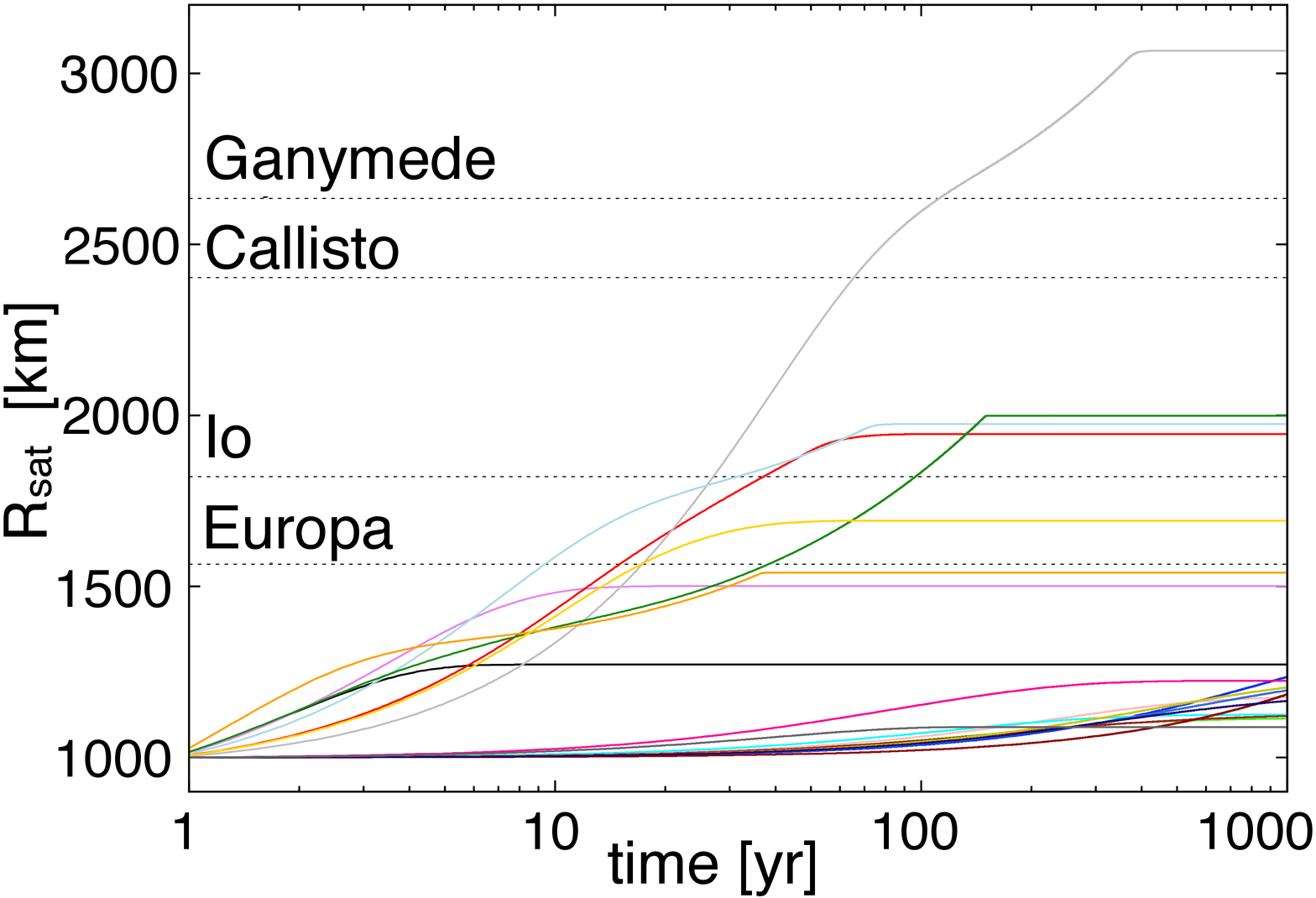}}\subfigure[]{\label{a-t-migII}\includegraphics[angle=90,width=.42\textwidth]{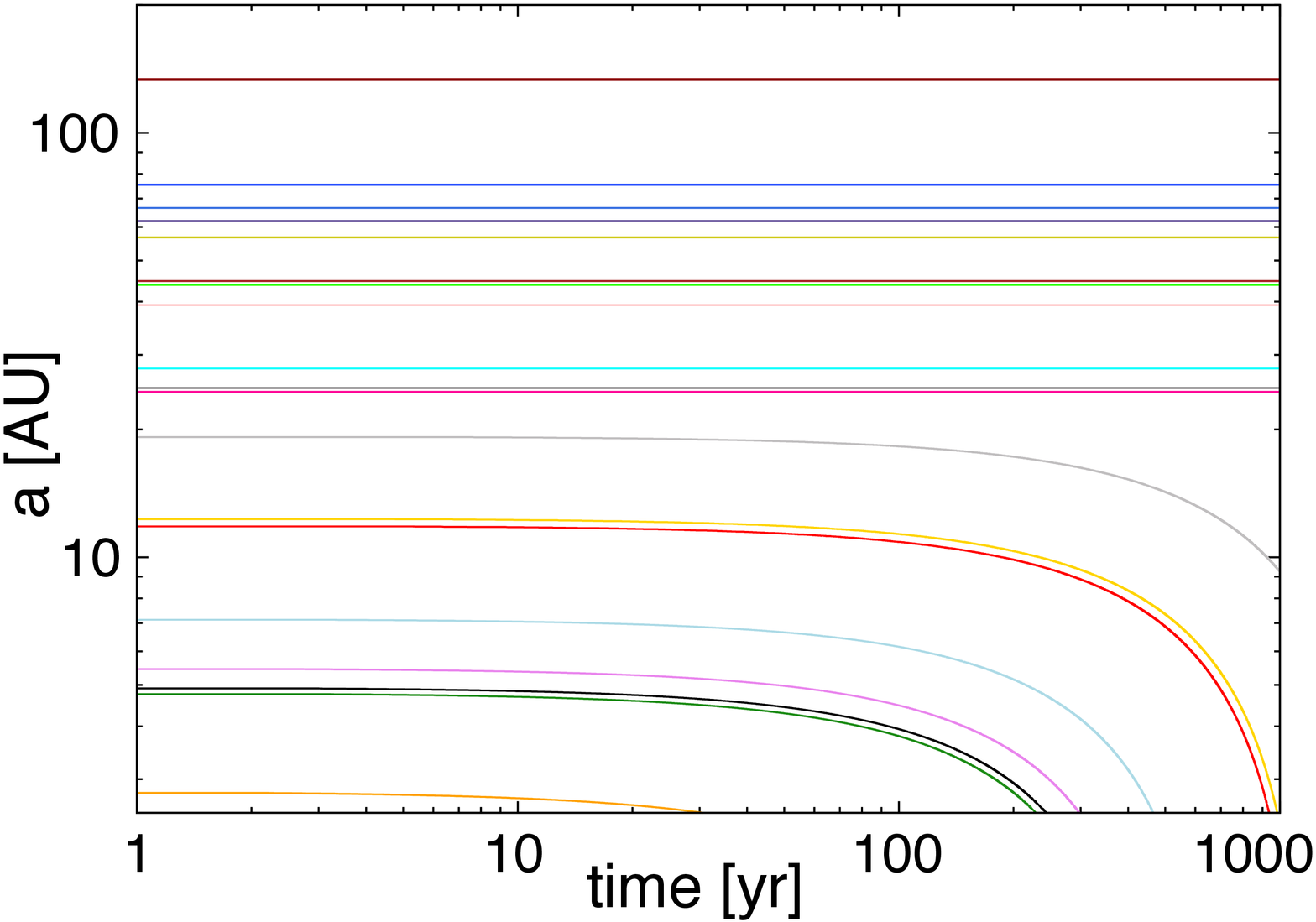}}
 \end{center}
  \caption{Evolution of radius (Figure \ref{r-t-migII}) and semimajor axis (Figure \ref{a-t-migII}) of satellites formed when including satellites migration.}
  \label{migII}
\end{figure*}

In this section we show simulation results when considering satellite-embryos orbital migration. Figure \ref{mamII} shows satellites' mass vs. semimajor axis (light blue dots). Satellites formed in situ (section \ref{sa}) are also shown with black open circles for comparison. Satellites migrate mostly in the fast disk-dominated type II regime and all the satellites initially located in the inner disk migrate towards Jupiter. Satellites plotted at the inner disk boundary are the ones that reached the disk edge and were removed. When these satellites reach the inner disk they already accreted all the material in their feeding zones, leaving no solids available for the formation of new generations of satellites. At the end of the $10^4$ years, satellites formed in the outer disk have never acquired the necessary mass to open up a gap and they do not migrate due to type II migration. Type I migration in such a low $\Sigma_g$ region is very slow and the bodies' migrations are not visible.
Figure \ref{migII} shows the evolution of satellite's radius (\ref{r-t-migII}) and semimajor axis (\ref{a-t-migII}).

\begin{figure}
\begin{center}
\includegraphics[angle=90,scale=.28]{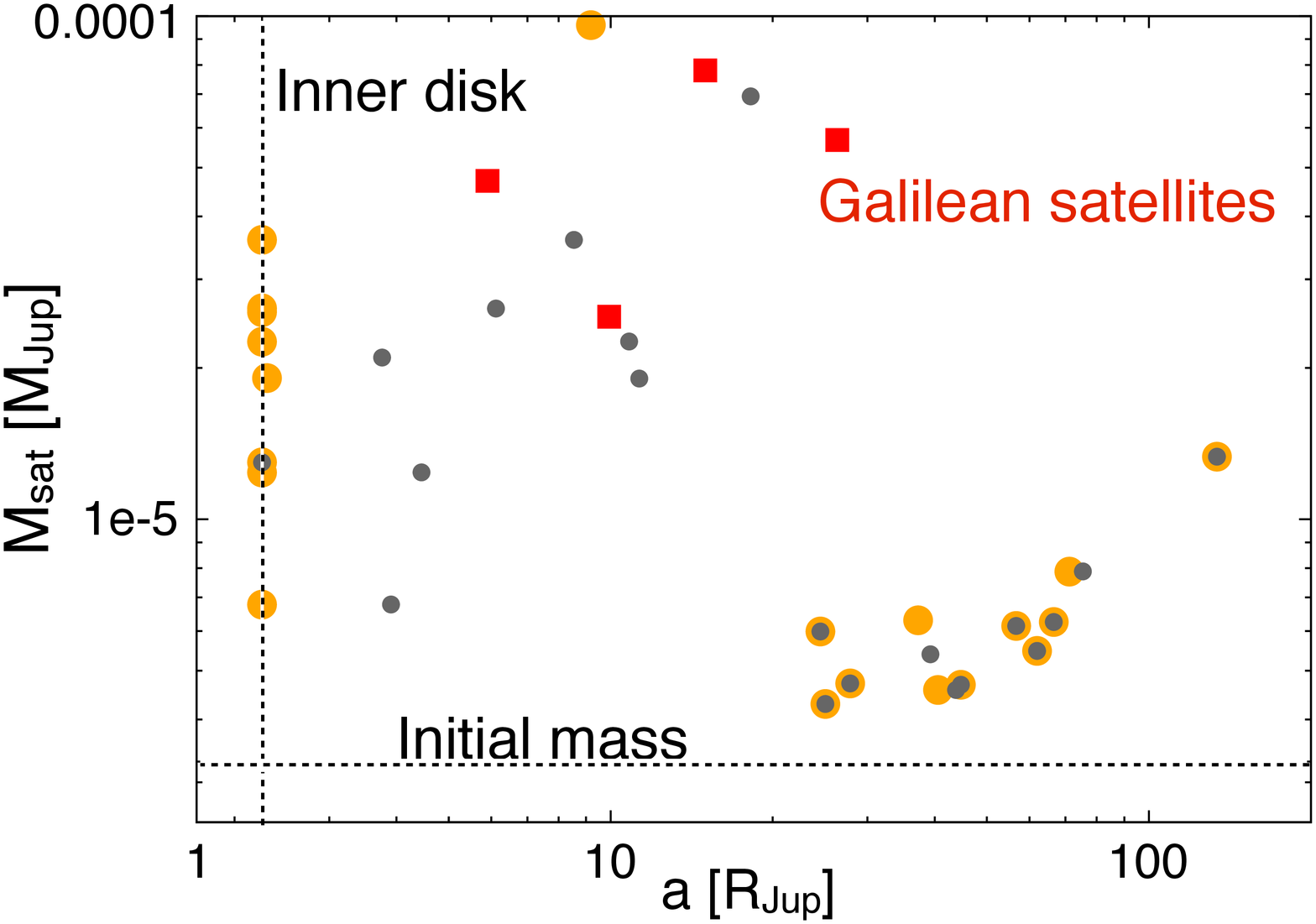}
\caption{Mass vs. semi major axis of satellites formed when considering satellites migration delayed by a factor $C_{migII}$. Orange dots are satellites formed with $C_{migII}=0.1$ and grey dots are formed with $C_{migII}=0.01$.}
\label{mamII-2}
\end{center}
\end{figure}

\paragraph{Slowing down type II migration} 
Since satellites migrate very fast and fall into Jupiter, we explore the results when slowing down type II migration by 10 and 100 times. Figure \ref{mamII-2} shows the mass and semimajor axis of satellites formed adopting $C_{migII}=0.1$ (orange) and $C_{migII}=0.01$ (grey dots). The simulations performed with $C_{migII}=0.01$ are the ones that favors the formation of Galilean Satellites.

\subsubsection{Satellitesimals growth with all the effects included}\label{sectionall}

\begin{figure}
\begin{center}
\includegraphics[angle=90,scale=.28]{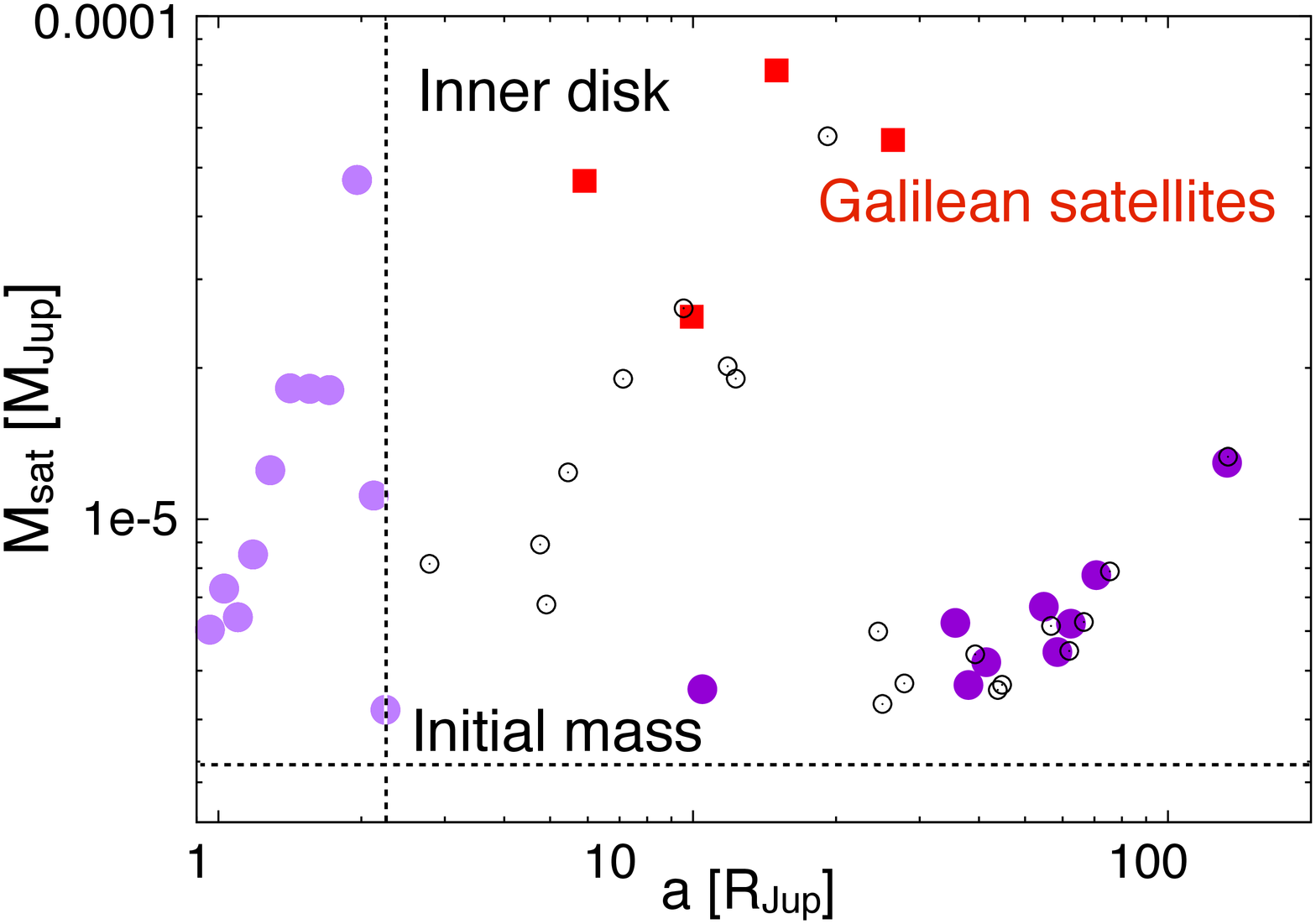}
\caption{Mass vs. semi major axis of satellites formed with all the effects included (purple dots). Lighter purple dots are satellites pushed inside the inner disk cavity. Black open circles are the case of growth in situ and without satellitesimals drag. Red squares are Galilean Satellites.}
\label{maall0}
\end{center}
\end{figure}

In this section we show the results of one satellite system formed when considering resonance trapping, satellites migration, and solids and gaseous disk evolution. 
The result is shown in Figure \ref{maall0}. 

Although satellites inside the disk inner edge do not suffer type II migration, the satellites are pushed inward by the resonant perturbations from satellites in the disk that are losing angular momentum by type II migration. As a result, almost all satellites are inside the disk inner edge. If the inner edge coincides with the corotation radius, they would eventually tidally decay to the host planet and the final satellite distribution is similar to that in \ref{mamII}.

N-body simulations by \citet{og12} reproduced systems in Laplace resonant state like Galilean satellite system, by taking into account "edge torque" exerted at the disk edge \citep{og10} due to dynamical friction that prevents the satellite system from falling onto the planet. However, in the low-viscosity disks that we are considering, since a gap is easy to be opened up, dynamical friction is not effective.Therefore, we cannot have reproduced a close-in resonant satellite system.

\begin{figure}
\begin{center}
\includegraphics[angle=90,scale=.28]{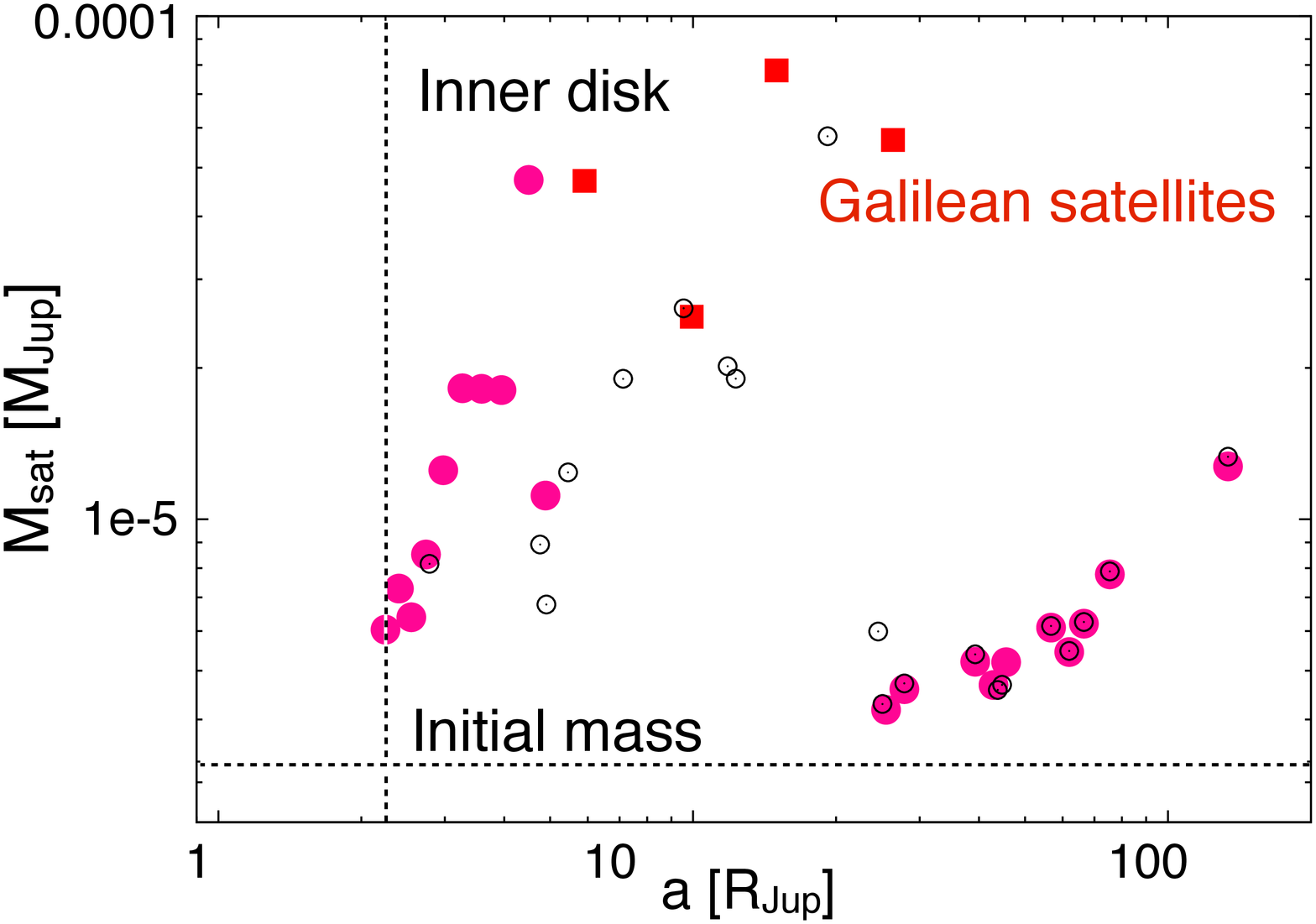}
\caption{Mass vs. semi major axis of satellites formed with all the effects included when considering the edge torque as a limiting case (pink dots). The case with growth in situ is shown as a comparison (open circles).}
\label{maall}
\end{center}
\end{figure}

Because the dynamics near the disk edge is not clear enough, we also carried out simulations with the edge torque as a limiting case, and the result is shown in Figure \ref{maall}. Even using the fastest type II migration regime ($C_{migII}=1$), most satellites do not reach the inner boundary. When migrating, all satellites in the inner region come too close to each other. The one closer to Jupiter reaches the inner disk. When the second closer to Jupiter approaches, they enter in a resonance that prevents the second satellite to keep migrating. This process continues until all the satellites formed in the inner disk are trapped in resonances with each other and can not move anymore. Note that a simulation longer than $10^4$ years does not change the results because satellites in the inner disk have no more solids to grow and they can not migrate because they are trapped in resonances. Increasing the dissipation time-scale affects only satellites formed in the outer disk, giving them more time to grow and migrate. However, although satellites remain, they are far smaller than Galilean satellites, because satellitesimals are rapidly depleted by gas drag, in particular in inner regions as shown in Figure \ref{disc-evolution-gas100}.

\subsection{Population Synthesis}\label{pop}
\paragraph{Parameters adopted in the simulations} \label{parameters-pop}

So far we have shown the results of single systems. In this section, we show the results of population synthesis simulations by superposing the results of 100 systems with different initial locations of satellite seeds. These simulations are calculated with all the effects taken into account (resonance trapping, satellites migration, and solids and gaseous disk evolution), which allows us to explore different parameters and investigate the most favorable conditions for the formation of Jovian satellites in a low-viscosity disk. 

As explained in section \ref{parameters}, the formation of each satellite system starts with 20 satellite-embryos of 1000 km in size located randomly in the disk. We also performed runs starting from only 4 embryos, corresponding to the number of Galilean satellites and find that varying the initial number of seeds doesn't change the results significantly. We allow the formation of a new generation of satellites in our simulations. To represent low-viscosity-disks, each satellite system has an $\alpha$ parameter taken randomly between $10^{-6}$ and $10^{-4}$ \citep{mea03a}.

In each simulation of 100 systems, we vary the size of the satellitesimals in the solid disk, adopting satellitesimals of 1, 10, 20 and 30 km, exploring the relevance of these sizes in satellites formation (section \ref{sat}). In some simulations we take the dissipation time-scale for each disk from a log-uniform distribution between $10^4$ and $10^7$ years \citep{mea03a,fu14}, but we also perform simulations where all satellite systems have the same dissipation time-scales of $10^4$, $10^5$ and $10^6$ years to see the differences in the populations formed (section \ref{dt}). We perform simulations using a disk with a higher concentration of solids (section \ref{section-gas}) and different migration rates (sections \ref{section-migII} and \ref{gas-migII}).

\subsubsection{The effect of dissipation time-scale}\label{dt}

\begin{figure*}
  \begin{center}
\subfigure[]{\label{104}\includegraphics[angle=90,width=.42\textwidth]{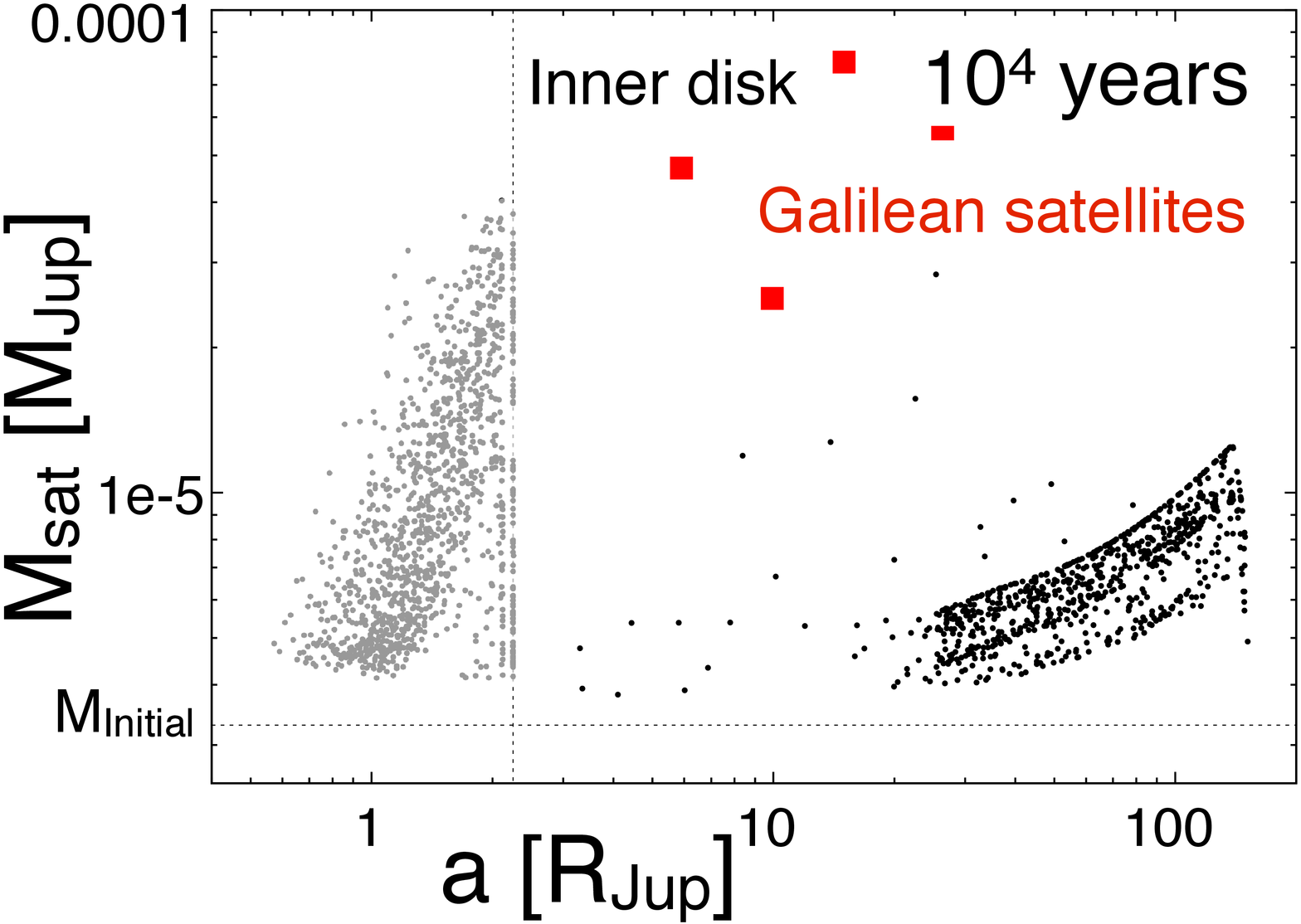}}\subfigure[]{\label{105}\includegraphics[angle=90,width=.42\textwidth]{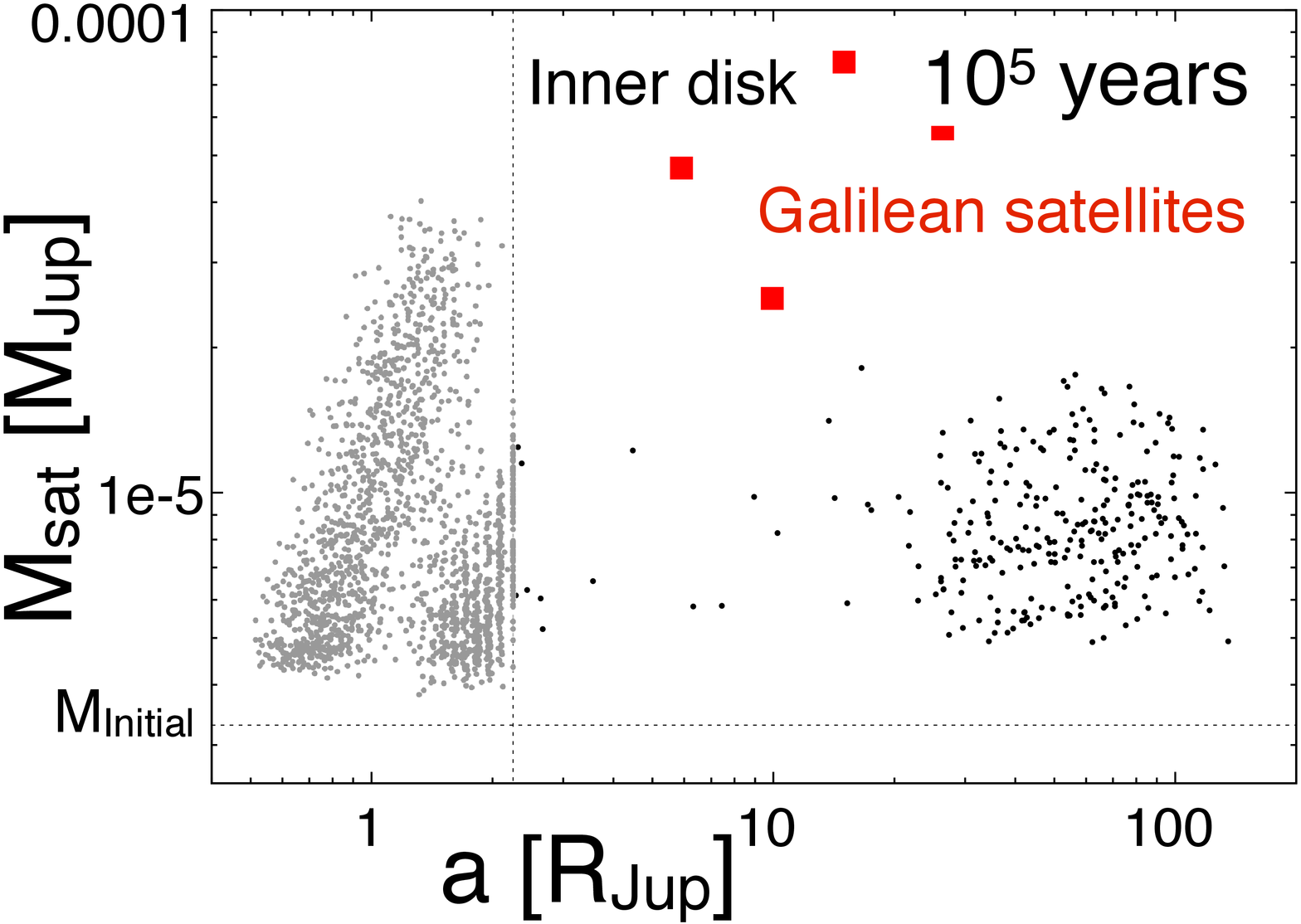}}

\subfigure[]{\label{106}\includegraphics[angle=90,width=.42\textwidth]{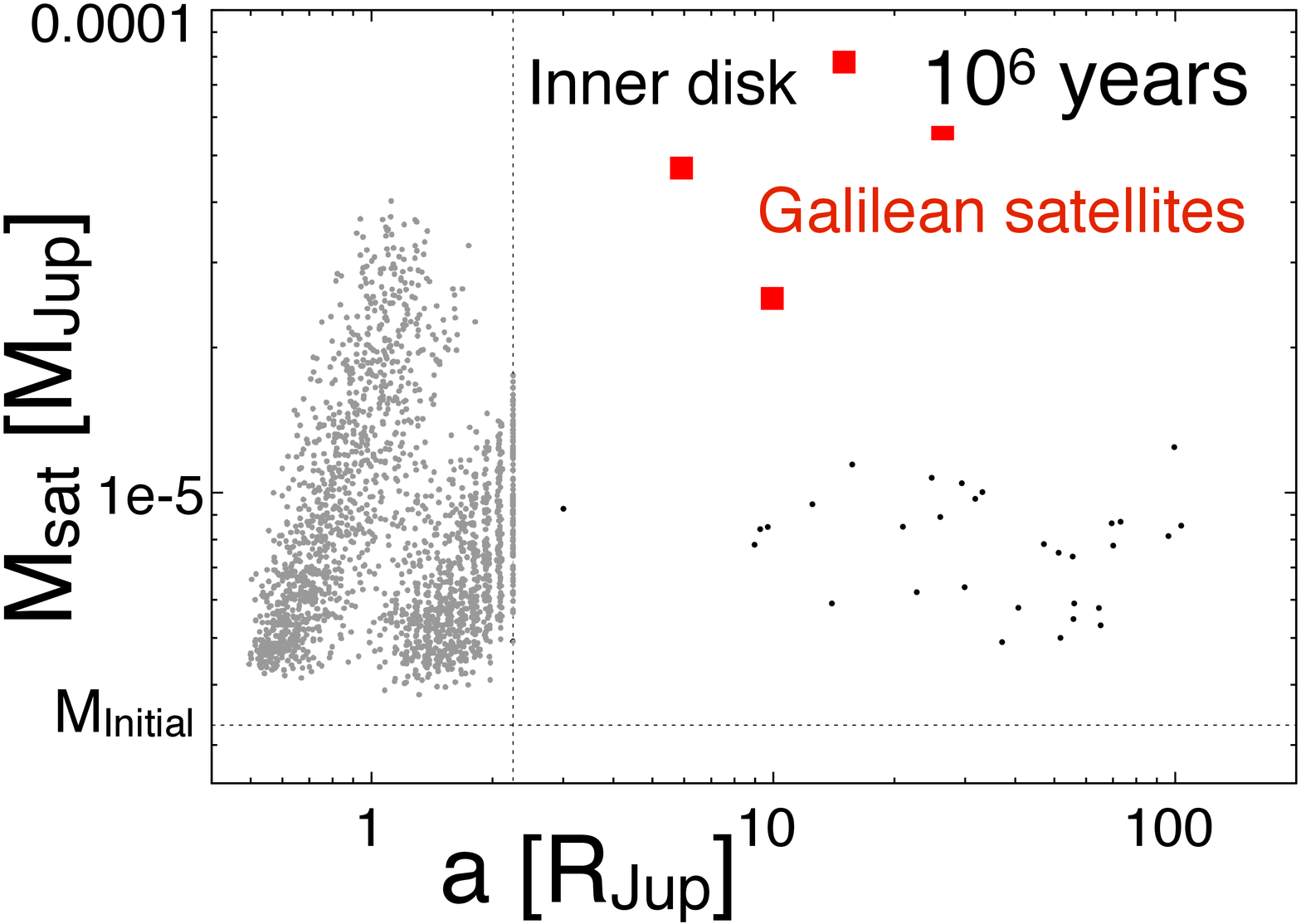}}
 \end{center}
  \caption{Mass and semimajor axis of all the satellites formed in 100 satellites systems when adopting different dissipation time-scales: $10^4$ (Fig. \ref{104}), $10^5$ (Fig. \ref{105}) and $10^6$ years (Fig. \ref{106}). Surviving satellites are black dots, grey dots are satellites inside the disk cavity.}
  \label{tau_disk}
\end{figure*}

Following \citet{fu14}, \citet{tu14} and \citet{mea03a}, we model a long-lasting disk with very low viscosity. Figure \ref{tau_disk} shows the population of satellites formed when using 30 km as the satellitesimals size and adopting $10^4$ (Fig. \ref{104}), $10^5$ (Fig. \ref{105}) and $10^6$ years (Fig. \ref{106}) as the lifetime of the gaseous disk.
As seen in section \ref{sectionall} for single systems, the bodies in the inner region are trapped in resonances and do not evolve much once they accreted all the solids available. Therefore, the main differences between the simulations with different time-scales is in the evolution of the satellites formed in the outer disk. This external part of the solid disk takes longer to dissipate (see Figure \ref{disc-evolution-gas100} ), giving the satellites more time to grow and evolve. In the simulations with long $\tau_{disk}$, outer satellites grow and start to migrate (Figure \ref{105}) and even reach the inner disk when they have more time, until they are trapped in a resonance. Figure \ref{106} shows that many satellites are in the inner region and much less remained in the outer parts, which makes more difficult the formation of satellites such as Ganymede and Callisto. 

\subsubsection{The effect of satellitesimals' size}\label{sat}

\begin{figure*}
  \begin{center}
\subfigure[]{\label{1km}\includegraphics[angle=90,width=.42\textwidth]{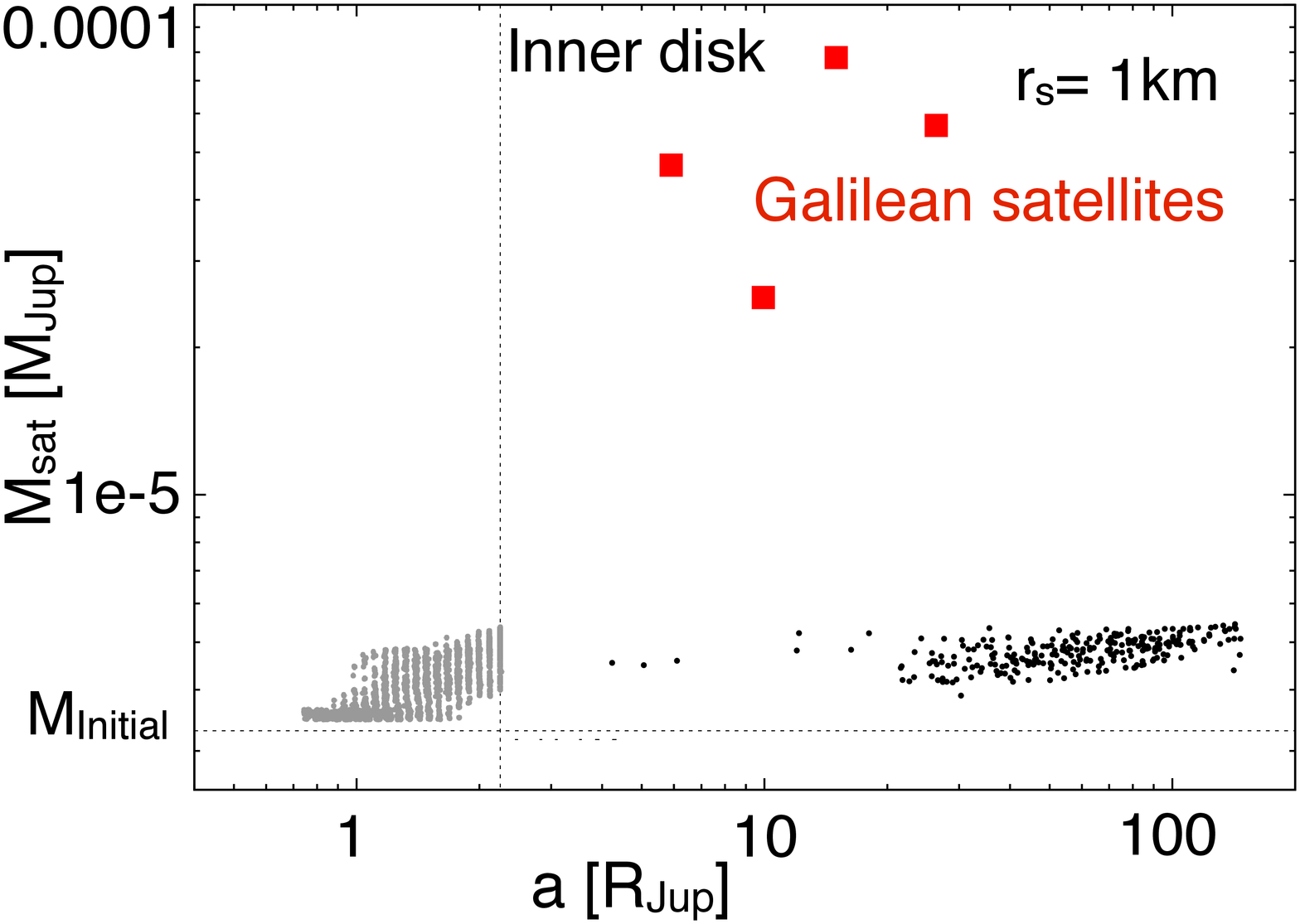}}\subfigure[]{\label{10km}\includegraphics[angle=90,width=.42\textwidth]{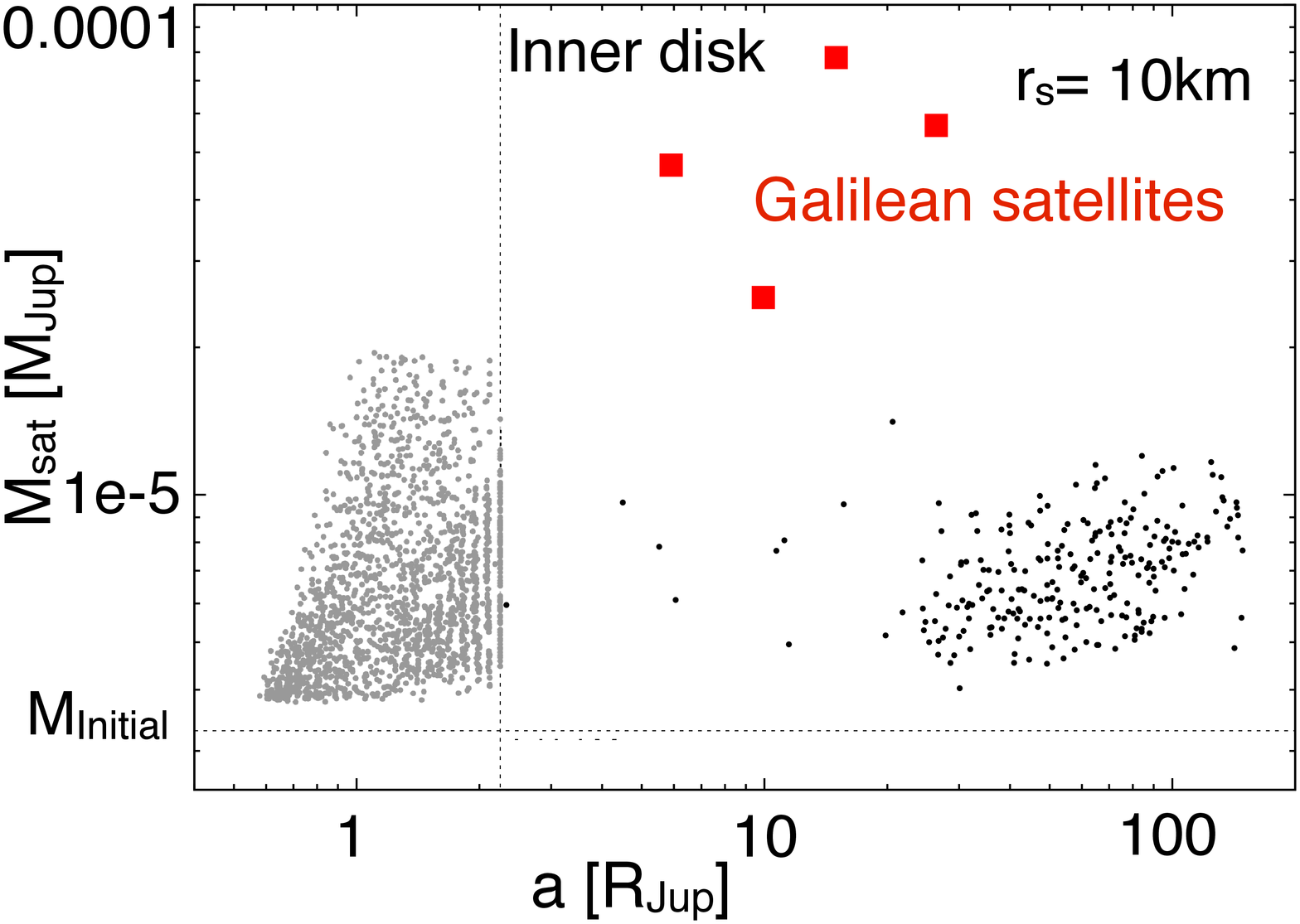}}
\subfigure[]{\label{20km}\includegraphics[angle=90,width=.42\textwidth]{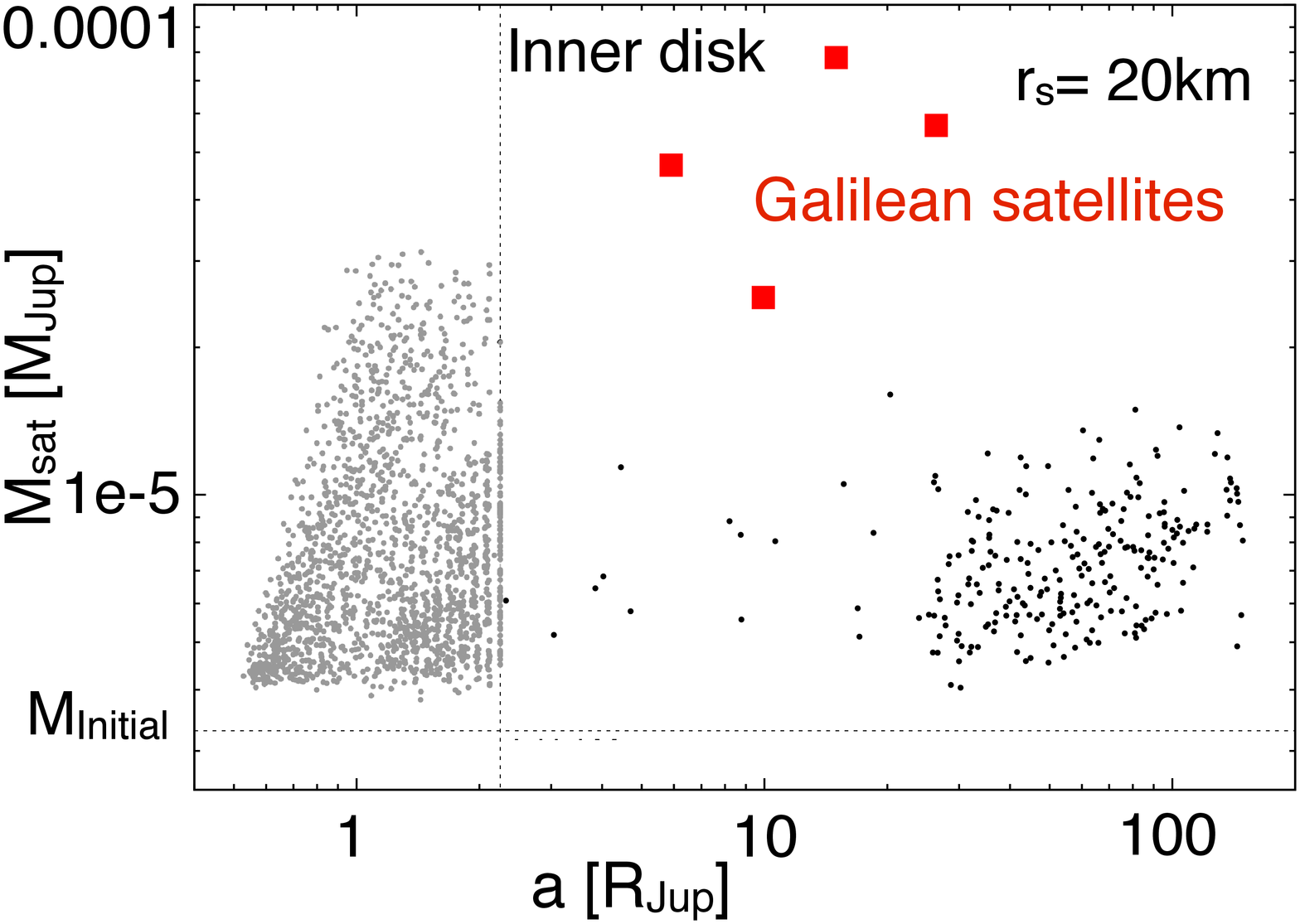}}\subfigure[]{\label{30km}\includegraphics[angle=90,width=.42\textwidth]{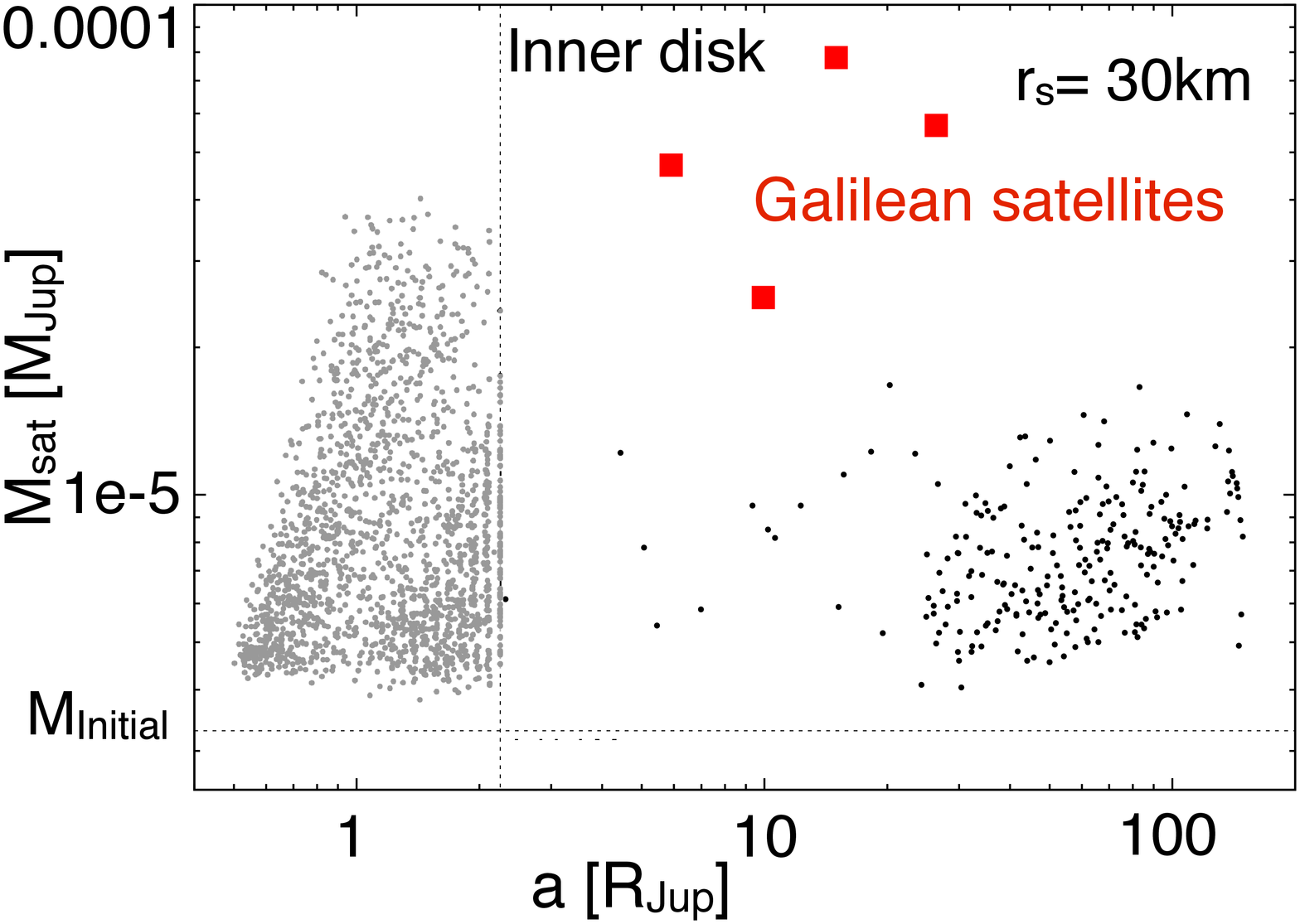}}
 \end{center}
  \caption{Population generated when taking $\tau_{disk}$ random from a log-uniform distribution between $10^4$ and $10^7$ yr and varying the radius of the satellitesimals in the solid disk. We use 1 km (Figure \ref{1km}), 10 km (Figure \ref{10km}), 20 km (Figure \ref{20km}) and 30 km of radius. The differences are due to the different lifetimes of the solid disk in each case. }
  \label{makm}
\end{figure*}

Since the size of the satellitesimals in the disk is extremely important (see Figure \ref{disc-evolution-gas100}), we explore the population of satellites formed when taking different radius for the satellitesimals that compose the solid disk, which translates into different solid disk lifetimes (see section \ref{nebula-evolution}). For these simulations, we take the dissipation time-scale random from a log-uniform distribution (section \ref{parameters-pop}). Figure \ref{makm} shows the population when using satellitesimals of 1 km (Figure \ref{1km}), 10 km (Figure \ref{10km}), 20 km (Figure \ref{20km}) and 30 km of radius (Figure \ref{30km}). The satellitesimals' size changes the lifetime of the solid disks and therefore, the final mass of the satellites is highly affected. 

In the population formed with satellitesimals of 1 km (Figure \ref{1km}) the most massive satellite hardly reaches at most $2 \times10^{-5}~M_{Jup}$. Disks with satellitesimals of 1km do not explain the formation of any Galilean satellite. When going to larger satellitesimal sizes, the disk of solids stays for longer time and the satellites reach larger masses. Some bodies generated when using 30 km (Fig. \ref{30km}) have masses similar to Io's and Europa's.
However, even in this case, the large satellites have migrated to inside of the disk inner edge. If the inner cavity size is much larger, satellites as massive as Io or Europa could be formed in larger orbital radius. But it is very unlikely that satellites as massive as Ganymede and Callisto are formed in outer regions.

\subsubsection{The effect of type II migration} \label{section-migII} 
 
\begin{figure*}
  \begin{center}
\subfigure[]{\label{cmigII1}\includegraphics[angle=90,width=.42\textwidth]{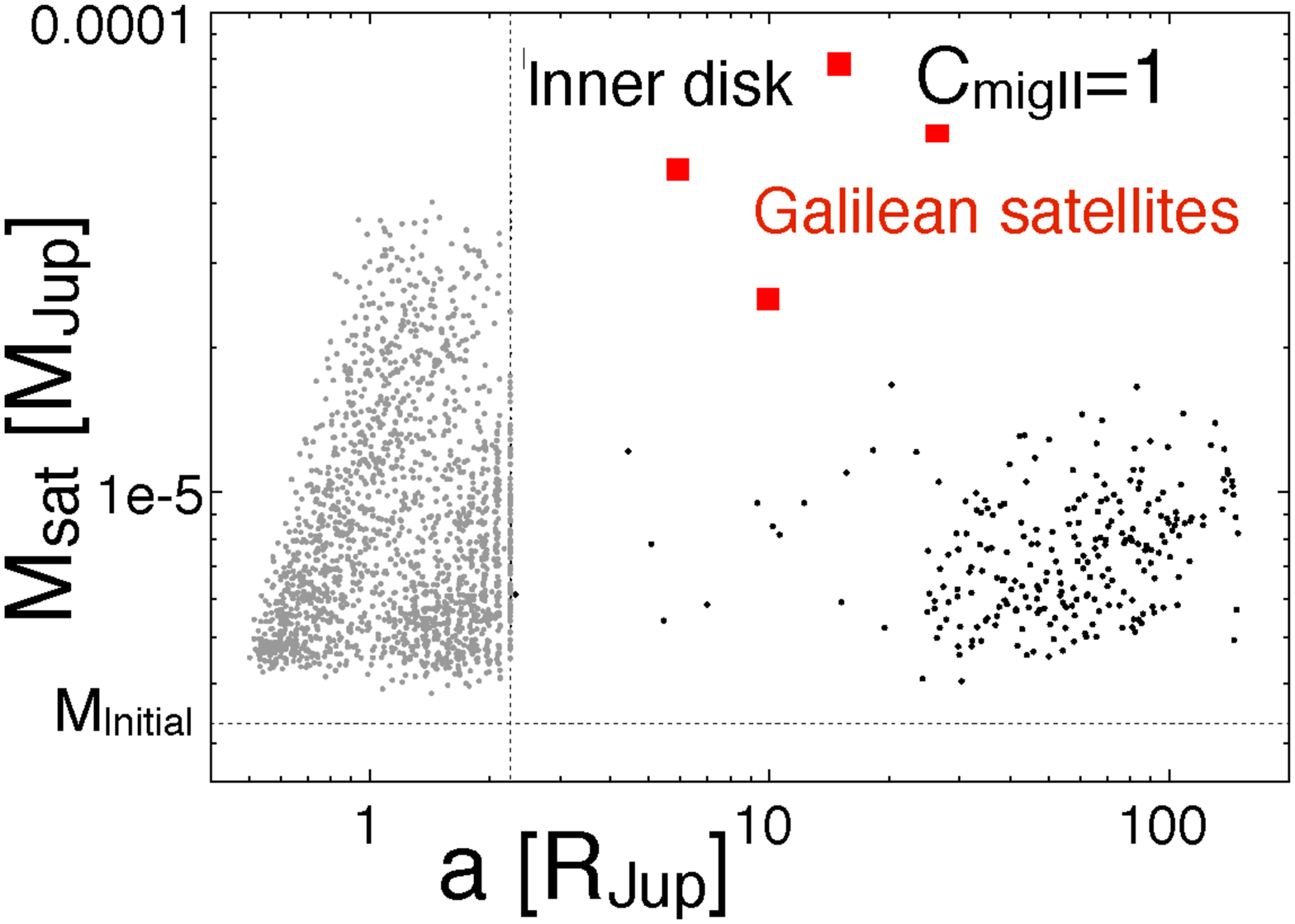}}\subfigure[]{\label{cmigII01}\includegraphics[angle=90,width=.42\textwidth]{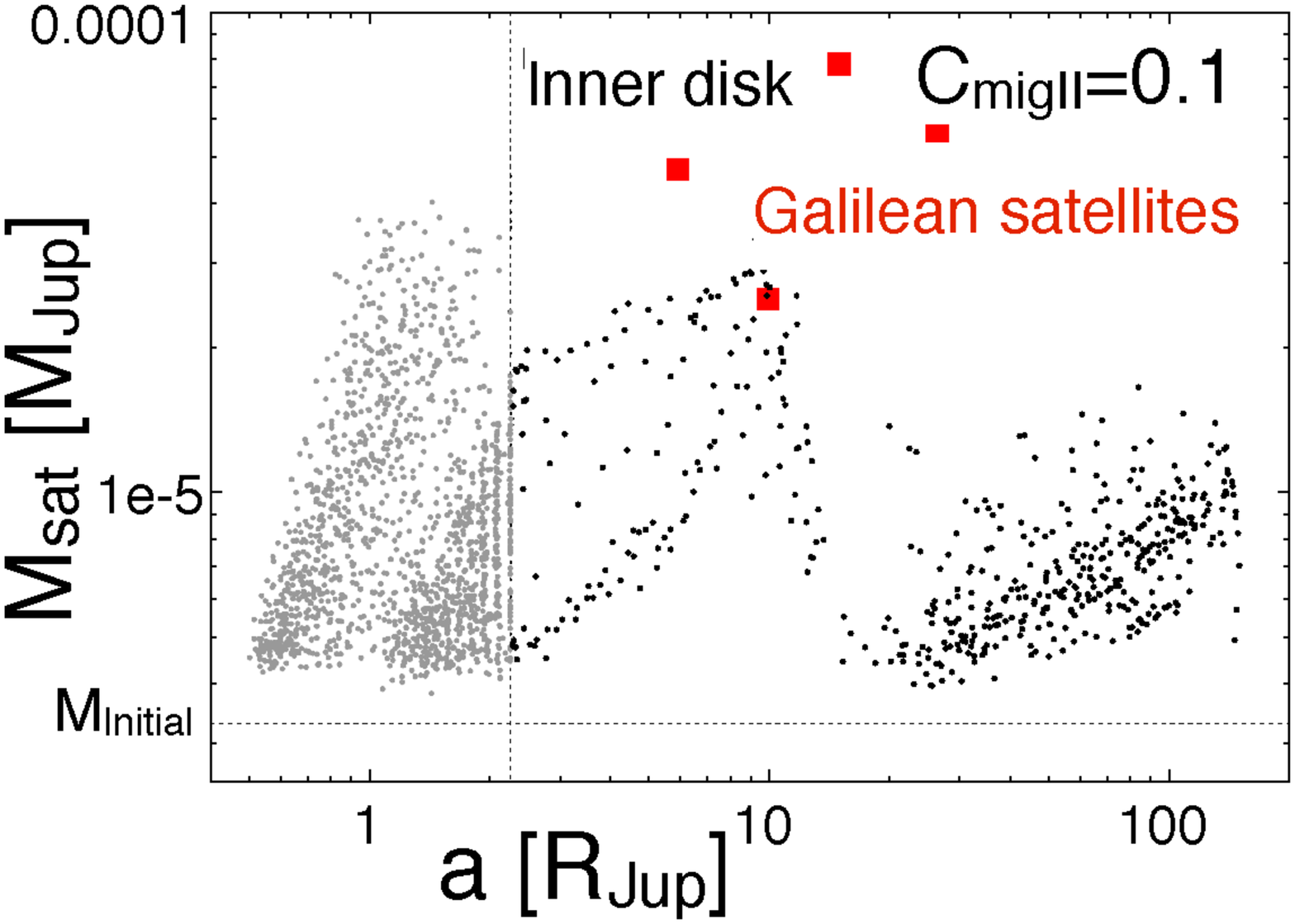}}
\subfigure[]{\label{cmigII001}

\includegraphics[angle=90,width=.42\textwidth]{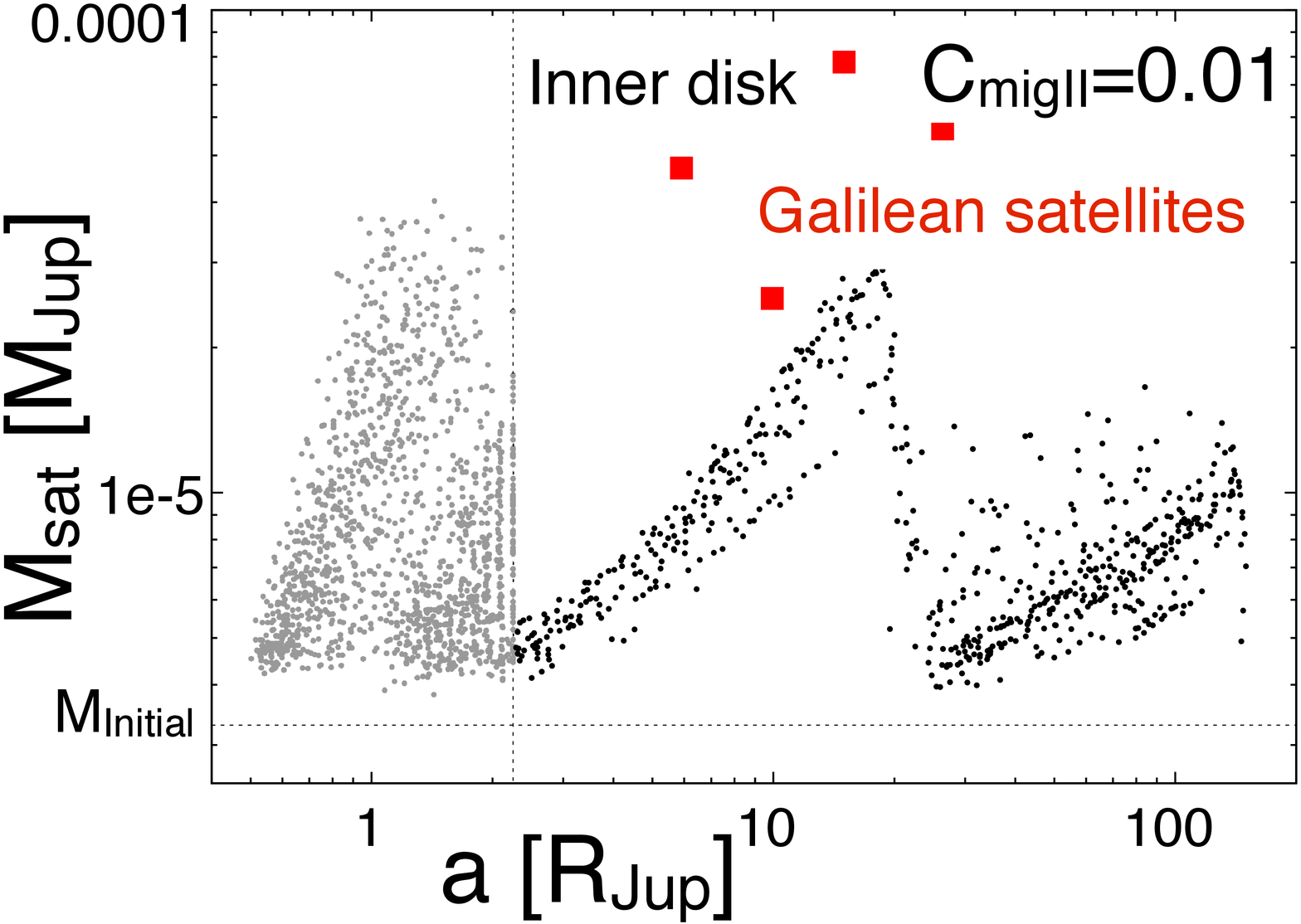}}
 \end{center}
  \caption{Mass vs. semimajor axis of satellites formed using different type II migration rates: $C_{migII}=1$ (Figure \ref{cmigII1}) , $C_{migII}=0.1$ (Figure \ref{cmigII01}) and $C_{migII}=0.01$ (Figure \ref{cmigII001}).}
  \label{cmigII}
\end{figure*}

In this section we explore the effects of different type II migration rates on the distribution of satellites formed. Figure \ref{cmigII} shows mass and semimajor axis of satellites formed considering satellitesimals of 30 km, random dissipation time-scale and adopting different type II migration time-scales. 

 When using $C_{migII}=1$ (Figure \ref{cmigII1}) there are very few satellites formed inside 20 R$_{Jup}$, and all of them with masses much lower than the Galilean satellites. When using a slower migration rate such as $C_{migII}=0.01$ (Figure \ref{cmigII1}), satellites in this region are formed, which favors the formation of satellites further out from Jupiter. Nevertheless satellites formed between 10 and 20 R$_{Jup}$ do not reach the masses of Ganymede and Callisto. A larger concentration of solids is needed. 

\subsubsection{The effect of heavy elements enhancement in the nebula} \label{section-gas} 

Precise measurements from the Galileo probe suggest that Jupiter has a hydrogen and helium envelope which is enriched in heavy elements compared to the solar composition \citep{ow99,at99,ma00}. In order to take this measurements into account, we performed simulations with a higher concentration of solids in the disk. In this section we show the results when assuming a gas to dust ratio of 10, which can be reached in two different ways: 
\begin{itemize}
\item [I] Increasing the amount of solids in the disk, but maintaining the same gas surface density. This can be achieved by considering that heliocentric planetesimals that cross Jupiter circumplanetary disk are captured, increasing the concentration of solids in the disk. 
\item [II] Decreasing the gas surface density and maintaining the solids in the disk. Which implies that some of the gas already dissipated when the formation of Galilean satellites started.  
\end{itemize}

\begin{figure*}
  \begin{center}
\subfigure[]{\label{10gas}\includegraphics[angle=90,width=.42\textwidth]{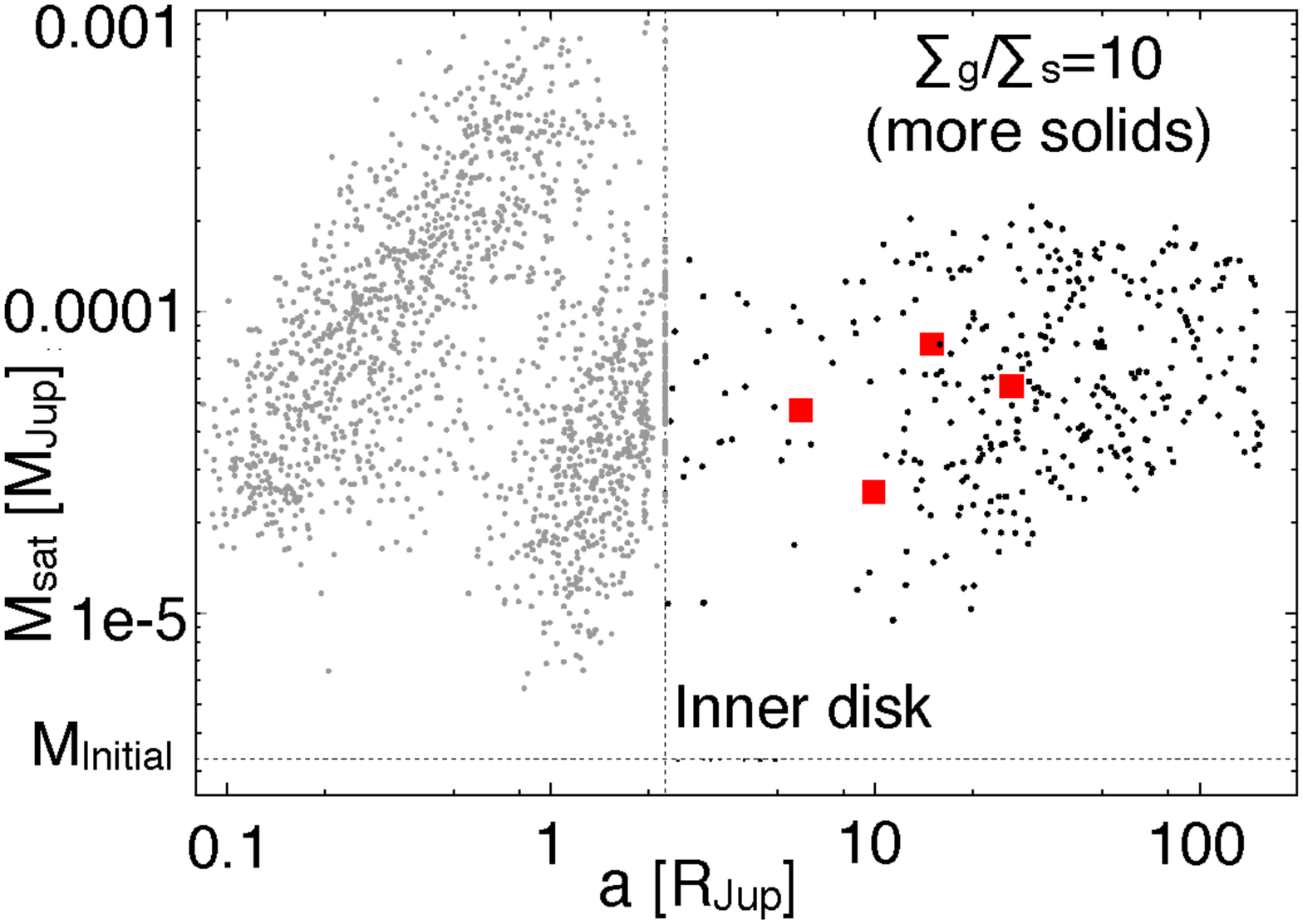}}
\subfigure[]{\label{10solids}\includegraphics[angle=90,width=.42\textwidth]{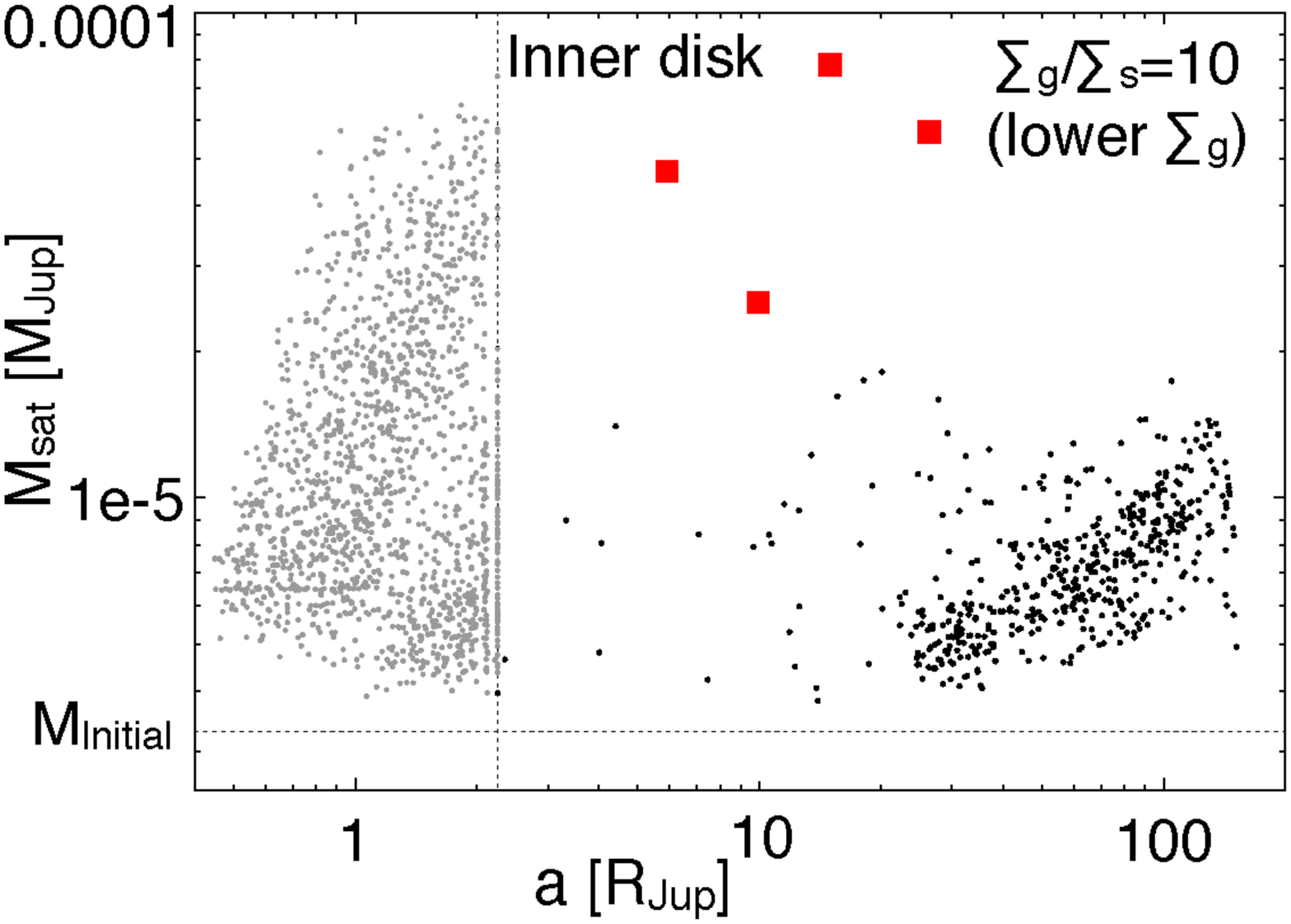}}
 \end{center}
  \caption{Mass and semimajor axis of satellites formed when adopting a $\frac{\Sigma_g}{\Sigma_s}=10$. In Figure \ref{10gas} we increase the solids and keep the same amount of gas in the disk, while in Figure \ref{10solids} we decreased the gas and kept the same solids surface density as in the previous cases.}
  \label{10disk}
\end{figure*}

When adopting I in our simulations (Figure \ref{10gas}), we find $\sim 20\%$ of surviving satellites that do not fall into the cavity. Some of these satellites have masses comparable to Galilean satellites. Nevertheless, we also find some satellites much larger than the ones observed, consequence of not considering a minimum mass disk model any more. 

On the other hand, when adopting II in the simulations (Figure \ref{10solids}), we are still considering a minimum mass disk model for solids but with a much larger concentration of solids relative to gas. The lower amount of gas in these simulations makes the migration rates slower which increases the chances of forming the Jovian satellites. Although, we are still unable to form satellites with similar masses and orbital radii to those of the four Galilean satellites with this distribution.  

\paragraph{The effect of an enhancement in heavy elements in the nebula + different migration rates} \label{gas-migII} 

\begin{figure}
  \begin{center}
\subfigure[]{\label{10solids-30km-cmigII01}\includegraphics[angle=90,width=.42\textwidth]{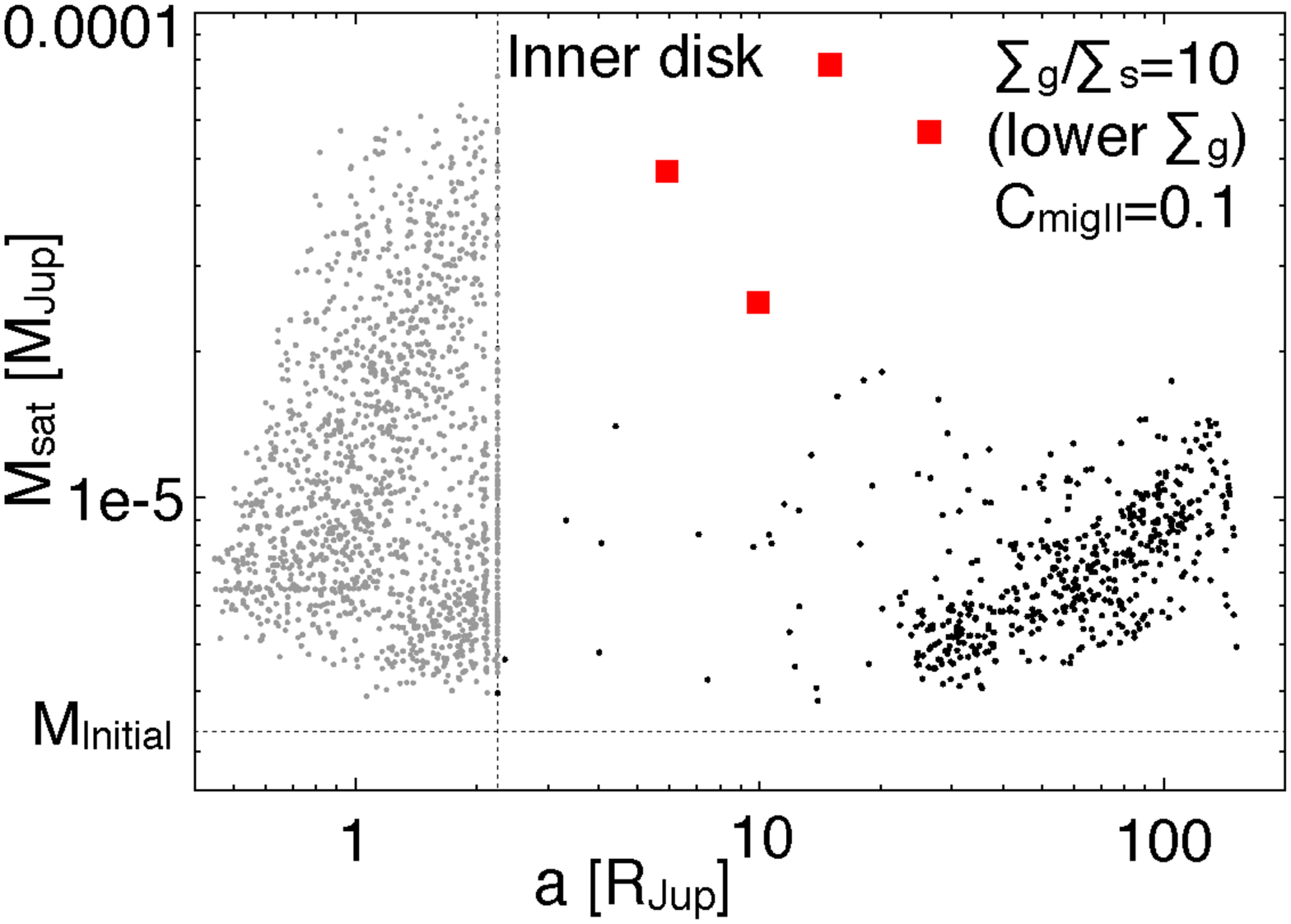}}

\subfigure[]{\label{10solids-30km-cmigII001}\includegraphics[angle=90,width=.42\textwidth]{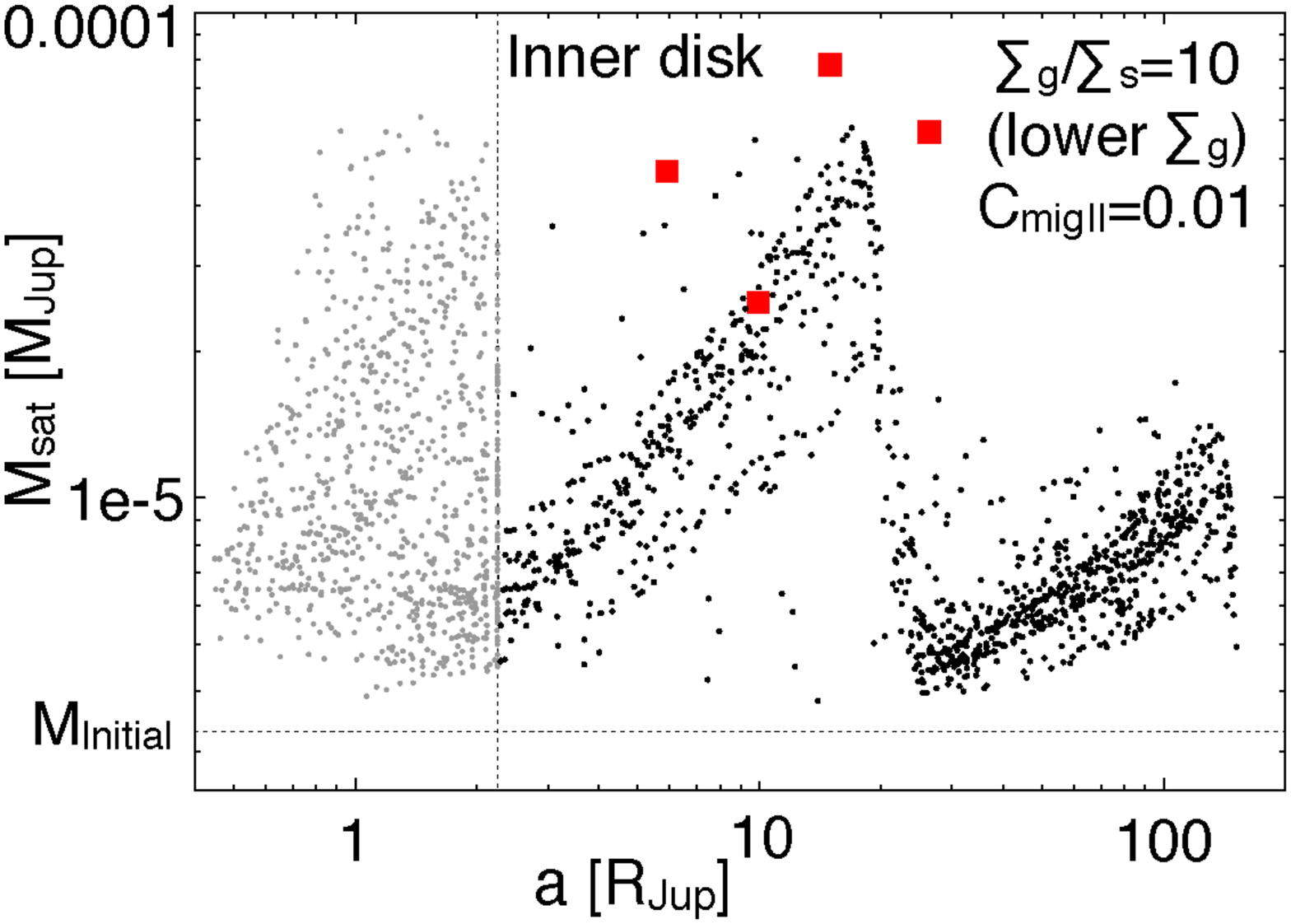}}
 \end{center}
  \caption{Mass vs. semimajor axis of the population of satellites formed when the gas surface density is decreased and the solids are maintained in order to have a gas to dust ratio of 10. Different migration rates are considered: $C_{migII}=0.1$ (Figure \ref{10solids-30km-cmigII01}) and $C_{migII}=0.01$ (Figure \ref{10solids-30km-cmigII001}).}
  \label{10solids-cmigII}
\end{figure}

In Figure \ref{10solids} we use $C_{migII}=1$ for establishing the migration rate. We also performed simulations using condition II and $C_{migII}=0.1$ (Figure \ref{10solids-30km-cmigII001} ) and $C_{migII}=0.01$ (Figure \ref{10solids-30km-cmigII001}). Our results show that condition II and $C_{migII}=0.01$ (Figure \ref{10solids-30km-cmigII001}) present the most favorable conditions for the formation of Galilean satellites. We note that even in this favorable case we are unable to form satellites as massive as Ganymede or as massive as Callisto and located in outer regions. One possibility to increase the probabilities of forming Ganymede and Callisto would be to have a larger inner cavity in the disk. This will shift resonant satellites outwards allowing the formation of satellites located further out in the disk. Giving the uncertainties in the determination of the inner disk boundary (see section \ref{disc-1}) this will be explored in a future work.

\section{Discussion: two different nebula models}

During Jupiter gas accretion and prior to gap opening, the turbulence in the protoplanetary disk generated by the inflow of material was too high, leading to extremely high temperatures that would prevent the ices to condense and the satellites to form \citep{ma99,kk06,e09}. Therefore formation of Galilean satellites most likely occurred after Jupiter opened up a gap and the inflow of gas in the disk decreased considerably. These studies, in addition to current MRI calculations \citep{fu14,tu14} favor the hypothesis of a low viscosity massive disk rather than an actively supplied high viscosity Jupiter nebula. 

However, semi-analytical studies performed by \citet{sa10} based on the actively supplied, gas-starved disk model, show a high success when forming the Jovian system. They found in $\sim 80 \%$ of their simulations 4 or 5 large bodies with masses similar to those of the Galilean satellites. A comparison between the initial parameters of this model and the model used in this paper can be seen in table \ref{comparison}. Since multiple generations of satellites are formed and consequently fall into Jupiter in this scenario \citep{cw02}, they started their simulations at a last stage (at $t=2\times10^6$ years), to consider the formation of the last and final generation of satellites only. We notice that in this model with very high $\alpha$ the main source of heating in the disk is viscous heating, and after the gap opening they imposed an exponential decay of the global temperature in the disk, independent on the viscosity. This behavior of the temperature might be a source of uncertainty in this kind of models. A different prescription for the evolution of the temperature gradient can lead to differences in the location of the ice line \citep{he15a}, in the condensation of the small satellitesimals that will form the satellites and therefore in the satellite's ice content. Finally, due to the low gas surface density, the drift of satellitesimals is negligible, and the migration rates are low which increases the chances of formation and survival of the Galilean satellites in the system, as we can see in their results.

A massive, low viscosity gaseous nebula, while favored in protoplanetary disks simulations, may not be the best scenario for the formation of Galilean satellites. As we show in this paper, the large gas surface density leads to a fast orbital migration of satellites and makes the satellitesimals drift very efficient, therefore depleting the disk of solids in a very short time, preventing satellites as big as Ganymede to form. We also show that a solid enhanced or gas poor scenario increases the chances of forming the satellites, although the initial conditions for such disks remain unknown. 

The description of the nebula that gave origin to the Galilean satellites is far from being resolved. Future missions such as JUICE \citep{gr13} or Europa-Clipper will provide more data to disentangle between these two main scenarios, leading to a better reconstruction of the history of this system. For example, the very fast accretion of satellites in the massive, low viscosity gaseous nebula model may result in silicate-iron differentiation in addition to differentiation between silicate/iron and ice. If these future missions determine whether the silicate-iron is differentiated or not in Ganymede, it may constrain formation mechanism.

\begin{table}
\centering
\caption{Comparison of Jupiter's nebula initial parameters used in two different population synthesis calculations performed with the minimum mass model (this paper) and the gas-starved disk model (as in \citet{sa10}).}
\begin{tabular}{lccc}
\hline
\hline
Model & Max. $\Sigma_{gas} (\frac{g}{cm^2})$ & $\alpha$ & $\tau_{disk}$ (yr) \\
\hline
Gas-starved disk & $100$ & $10^{-2}$-$10^{-3}$ & (3-5) $\times10^6$\\
Minimum mass & $10^5$ & $10^{-4}$-$10^{-6}$ & $10^4$-$10^7$\\
\hline
\label{comparison}
\end{tabular}
\end{table}

\section{Conclusion}

We developed a semi-analytical model to study formation of Galilean satellites using a massive, low viscosity (nearly stationary) disk model, which may be consistent with recent studies of magnetorotational instability in Jupiter's circumplanetary disk \citep{fu14,tu14}. In this scenario, Jupiter's regular satellites formation starts once the gas infall to Jupiter stops and the formation lasts for disk diffusion timescales of $10^4$ to $10^7$ years. We performed population synthesis calculations to form 100 satellite systems and investigate how resonance trapping, accretion and migration of satellites, disk evolution and solid's concentration in Jupiter's circumplanetary disk affect the distribution of the satellites formed, getting a better understanding of the initial conditions that lead to the formation of Galilean satellites. 

We find that due to the high gas density, the population of small satellitesimals migrate very rapidly due to gas drag and the solid disk is quickly depleted. The inner solid disk is depleted only in 100 years for a population of satellitesimals of 1km. If satellitesimals of 30km are considered, the solid disk lifetime is 1000 years, giving more time to the satellites to grow and evolve. Taking this fast satellitesimals migration into account, we start our calculations from satellite-embryos of 1000 km size, assuming that the embryos are formed by very rapid pebble accretion. This initial condition is the most optimal one against the loss of bodies by orbital migration. 

Because of the low viscosity, satellites open up a gap and undergo type II migration. Due to the large disk mass, the migration occurs in the disk-dominated regime, which is the fastest mode. Satellites stop migrating when they reach the inner disk boundary, but can be pushed inside the inner disk cavity due to migration of resonant satellites and eventually fall into Jupiter. In order to slow migration and take into account possible changes in type II migration rates due to non-linear effects, we introduce a factor ($C_{migII}$) in our calculations. We explore different values for $C_{migII}$, getting longer migration time-scales, preventing satellites to migrate very fast and getting a higher percentage of surviving satellites. Longer migration time-scales allow the formation of satellites located further out from the disk, presenting better conditions for the formation of Jovian satellites. Artificial elongation of the migration time-scale of 100 times allows us the formation of satellites at the location of Ganymede and Callisto. Nevertheless, these satellites were originally formed in the outer disk and did not reach the masses of those Jovian satellites. 

Jupiter's atmosphere is enriched in solids compared to the solar nebula. To take this into account, we performed some simulations adopting a larger concentration of solids in Jupiter's circumplanetary disk, which can be reached either by increasing the amount of solids in the nebula or decreasing the gas surface density. A larger solid surface density allows the formation of satellites similar to the Galilean ones, but also leads to the formation of a population of much larger satellites that we do not observe today. On the other hand, decreasing the amount of gas helps to get lower migration rates and favors the formation of satellites at the location of Jovian satellites, although the masses of these satellites do not match the masses of the four Jovian satellites at their present locations. 

Galilean satellites formation is not a trivial process. Lower migration rates, a larger concentration of solids and big building blocks favor their formation, although the large mass of Ganymede and location of Callisto further away from Jupiter are not easy to reach. In this paper we  moved towards a better understanding of the conditions that lead to their formation, further studies as well as data provided by future missions (JUICE, Europa-Clipper) will improve our understanding on this system.

\section{Acknowledgments}
We thank Takanori Sasaki, S\'ebastien Charnoz and an anonymous referee for useful comments that improved our paper.

\end{document}